\documentclass[a4paper,11pt, final]{article}
\pdfoutput=1
\usepackage[utf8]{inputenc}
\usepackage{jcappub}
\usepackage{amsmath}

\usepackage[inline]{showlabels}

\usepackage{xspace}
\newcommand*{\ie}{i.e., }
\newcommand*{\eg}{e.g., }
\newcommand*{\fig}{figure\@\xspace}
\newcommand*{\Eq}{eq.\@\xspace}
\newcommand*{\Eqs}{eqs.\@\xspace}

\usepackage{xifthen}
\newcommand*\diff{\mathrm{d}} 
\newcommand*\ldiff[2][]{ \ifthenelse{\isempty{#1}}{ \diff #2}{\diff^#1#2} \,} 
\let\limitint\int 
\renewcommand{\int}{\limitint \!} 

\title{Higgs inflation in Einstein-Cartan gravity}
\author[a]{Mikhail Shaposhnikov,}
\author[a]{Andrey Shkerin,}
\author[a]{Inar Timiryasov,}
\author[a]{Sebastian Zell}
\affiliation[a]{Institue of Physics, Laboratory for Particle Physics and Cosmology,\\
\'Ecole Polytechnique F\'ed\'erale de Lausanne, CH-1015 Lausanne, Switzerland}
\emailAdd{mikhail.shaposhnikov@epfl.ch}
\emailAdd{andrey.shkerin@epfl.ch}
\emailAdd{inar.timiryasov@epfl.ch}
\emailAdd{sebastian.zell@epfl.ch}

\abstract{

We study inflation driven by the Higgs field in the Einstein-Cartan formulation of gravity. In this theory, the presence of the Holst and Nieh-Yan terms with the Higgs field non-minimally coupled to them leads to three additional coupling constants. For a broad range of parameters, we find that inflation is both possible and consistent with observations. In most cases, the spectral index is given by $n_s=1-2/N_\star$ (with $N_\star$ the number of e-foldings) whereas the tensor-to-scalar ratio $r$ can vary between about $10^{-10}$ and $1$. Thus, there are scenarios of Higgs inflation in the Einstein-Cartan framework for which the detection of gravitational waves from inflation is possible in the near future. In certain limits, the known models of Higgs inflation in the metric and Palatini formulations of gravity are reproduced. Finally, we discuss the robustness of inflationary dynamics against quantum corrections due to the scalar and fermion fields.

}

\date{}

\begin{document}

\maketitle

\section{Introduction}
\label{sec:intro}

The last decades have witnessed a remarkable transition to the era of precision cosmology. An important example are measurements of the cosmic microwave background (CMB) with the constantly increasing accuracy \cite{BENNETT1993409,1001.4538,Akrami:2018odb}. Their results are fully compatible with the paradigm of inflation \cite{Starobinsky:1980te, Guth:1980zm, Linde:1981mu, Mukhanov:1981xt}. If the inflationary phase indeed took place in the early moments of our Universe, this leads to the question of the nature of inflaton field. An interesting idea is that the Higgs field can take on this role \cite{Bezrukov:2007ep}. This proposal stands out among numerous models of inflation in that it does not require degrees of freedom beyond those already present in the Standard Model and General Relativity.

Even if the inflaton is identified with the Higgs boson, this does not uniquely fix inflationary predictions. The reason is that Higgs inflation is sensitive to which formulation of gravity one uses. In the original proposal \cite{Bezrukov:2007ep}, the metric formulation was employed. In this theory, the connection $\Gamma^\rho_{\mu\nu}$ is a priori determined as a function of the metric $g_{\mu\nu}$. Thus, $g_{\mu\nu}$ is the only dynamical variable. Another version \cite{Bauer:2008zj} employs the Palatini variant of gravity, in which both $\Gamma^\rho_{\mu\nu}$ and $g_{\mu\nu}$ are treated as independent variables. Moreover, Higgs inflation in the teleparallel formulation of gravity has been studied recently \cite{Raatikainen:2019qey}. Finally, inflation driven by a non-minimally coupled scalar field has also been investigated in the framework of affine gravity \cite{Azri:2017uor}.

The metric and Palatini scenarios of Higgs inflation can be described by the same classical action.\footnote
{See \cite{Rubio:2018ogq} and \cite{Tenkanen:2020dge} for reviews of metric and Palatini Higgs inflation, respectively.}
Its gravitational part reads
\begin{equation} \label{shortAction}
S_{\rm grav} =  \int \ldiff[4]{x} \sqrt{-g} \left\{\frac{M_P^2+\xi h^2}{2}R  \right\}\;, 
\end{equation}
where $M_P$ is the reduced Planck mass and $h$ corresponds to the Higgs field in the unitary gauge. The parameter $\xi$ sets the strength of the non-minimal coupling between the Higgs field and gravity. In both versions of Higgs inflation, the requirement of matching the observed CMB normalization leads to $\xi\gg1$, although the order of magnitude of the non-minimal coupling is different in the two cases. 

The goal of the present paper is to generalize the previous studies of Higgs inflation by adopting the theory of gravity which encompasses both the metric and Palatini versions as special cases. For this purpose, we shall use the Einstein-Cartan (EC) framework \cite{Cartan:1922, Cartan:1923, Cartan:1924, Cartan1925, Einstein1928, Einstein19282, Utiyama:1956sy, sciama1962analogy, Kibble:1961ba}.\footnote
{See, e.g., \cite{Hehl:1976kj,Shapiro:2001rz} for reviews of EC theory.}
As in the Palatini theory, $\Gamma^\rho_{\mu\nu}$ and $g_{\mu\nu}$ are treated as independent variables, but the difference is that $\Gamma^\rho_{\mu\nu}$ is no longer assumed to be symmetric in the lower indices. This leads to the appearance of torsion:\footnote
{In the EC formalism, one still assumes that non-metricity is zero. We expect that this does not result in a loss of generality since our theory is invariant under projective transformations of the connection \cite{Dadhich:2010xa}, and one can use this invariance to eliminate non-metricity.}
\begin{equation}
T^\mu_{\;\;\nu\rho}=\Gamma^\mu_{\;\;\nu\rho}-\Gamma^\mu_{\;\;\rho\nu} \;,
\end{equation}
where $T^\mu_{\;\;\nu\rho}$ is the torsion tensor. It is important to note that torsion in EC gravity does not lead to new propagating degrees of freedom. Moreover, if one only considers the Higgs field and gravity with the action \eqref{shortAction}, it turns out that torsion vanishes dynamically, \ie the EC theory is equivalent to the Palatini formulation in this case.

Since in the EC formulation one does not assume \textit{a priori} that torsion vanishes, another difference to Palatini gravity arises. Namely, one can add non-trivial additional terms of mass dimension not greater than four to the action. We have already performed a systematic study of these terms in our companion paper \cite{Shaposhnikov:2020frq}. In the present paper, we only discuss the terms relevant for Higgs inflation. First, there is the so-called Holst action \cite{Hojman:1980kv, Nelson:1980ph, Castellani:1991et, Holst:1995pc}: 
\begin{equation}
\int \diff^4x\sqrt{-g}\:\epsilon^{\mu\nu\rho\sigma}R_{\mu\nu\rho\sigma} \;,
\end{equation}
where $R_{\mu\nu\rho\sigma}$ is the Riemann tensor as a function of the connection $\Gamma^\rho_{\mu\nu}$. Once $\Gamma^\rho_{\mu\nu}$ is no longer symmetric, $R_{\mu\nu\rho\sigma}$ also loses its symmetry properties so that the Holst action can give a non-trivial contribution. As was done for the Einstein-Hilbert term in \Eq \eqref{shortAction}, one can also add a non-minimal coupling of the Higgs field to the Holst term. Furthermore, in the presence of torsion another term of mass dimension two exists, namely the Nieh-Yan topological invariant \cite{Nieh:1981ww}
\begin{equation}
\int \diff^4x\, \partial_{\mu}\left(\sqrt{-g}\epsilon^{\mu\nu\rho\sigma}T_{\nu\rho\sigma}\right) \;. \label{Nieh-Yan}
\end{equation}
As the name indicates, the term \eqref{Nieh-Yan} only corresponds to a boundary term and therefore does not contribute to classical equations of motion. However, one can endow it with classical dynamics by introducing the non-minimal coupling of the Higgs field.\footnote
{There are two additional topological invariants, namely the Euler class  $\int\diff ^4x\sqrt{-g}\:\epsilon_{\mu\nu\rho\sigma}\epsilon^{\alpha\beta\gamma\delta}R^{\mu\nu}_{\;\;\;\;\alpha\beta}R^{\rho\sigma}_{\;\;\;\;\gamma\delta}$ and the Pontryagin class $\int \diff^4x\sqrt{-g}\:\epsilon^{\mu\nu\rho\sigma}R^{\alpha\beta}_{\;\;\;\;\mu\nu}R_{\alpha\beta\rho\sigma}$ (see, \eg \cite{Nieh:2013ada} for a summary). However, since they are already of mass dimension four, we do not consider a non-minimal coupling to them.}
To summarize, in the EC framework one can replace \Eq \eqref{shortAction} by the following more general gravitational action:
\begin{equation}\label{S_grav}
\begin{split}
S_{\rm grav} & =\frac{1}{2}\int \diff^4x\sqrt{-g}\:(M_P^2+\xi h^2)R \\
& + \frac{1}{4\bar{\gamma}}\int \diff^4x\sqrt{-g}\:(M_P^2+\xi_\gamma h^2)\epsilon^{\mu\nu\rho\sigma}R_{\mu\nu\rho\sigma}\\
& - \frac{1}{4}\int \diff^4x\:\xi_\eta h^2\partial_{\mu}\left(\sqrt{-g}\epsilon^{\mu\nu\rho\sigma}T_{\nu\rho\sigma}\right) \;.
\end{split}
\end{equation}
On top of the familiar parameter $\xi$, it contains three additional coupling constants $\bar{\gamma}$, $\xi_\gamma$ and $\xi_\eta$.

In the analysis of inflation, it is convenient to replace the theory \eqref{S_grav} by an equivalent theory in the metric formulation of gravity. It is obtained by, first, splitting the connection into its torsion-free and torsion-full parts, second, solving explicitly for the torsion-full part and, third, plugging the result back into the action; see \cite{Shaposhnikov:2020frq} for details of the computation. The resulting theory, in addition to the standard action of metric gravity, contains extra operators of mass dimension six. This way, one trades the additional terms of the EC formulation of gravity for a specific set of higher-dimensional operators. Of course, one could have started from the beginning with a metric theory supplemented by certain higher-dimensional operators, with the same result. However, it makes a difference from the point of view of effective field theory, since the original action of EC gravity only contains operators of dimension not higher than four.

Our interest in EC gravity is supported by the following observations. On the one hand, there is no compelling reason to exclude any consistent formulation of gravity, should that formulation be more general or lead to predictions different from the widely-used metric formulation. Our motivation to study EC gravity comes from the fact that it is a natural generalization of the Palatini gravity in the presence of fermions, which are present unavoidably in any theory comprising the Standard Model and gravity.\footnote
{Since the coupling of fermions to gravity relies on the spin connection, one should use the spin connection as dynamical variable instead of $\Gamma^\rho_{\mu\nu}$. This results in the EC formulation of gravity, which -- as stated -- is equivalent to the Palatini version in the absence of fermions.}
 On the other hand, EC gravity can be viewed as a gauge theory of the Lorentz group \cite{Kibble:1961ba}. This puts gravity on the same footing as the fundamental forces of the Standard Model, which is again favorable when one looks for a unification of the Standard Model with gravity.

A point of view that the Palatini formulation of gravity may be advantageous over the metric formulation has been elaborated on in \cite{Shaposhnikov:2020geh}, on the examples of Higgs inflation and the problem of hierarchy between the Higgs and Planck masses. With regard to inflation, the argument relies on the fact that the cutoff scale, above which perturbation theory breaks down, is higher in the Palatini model \cite{Bauer:2010jg} than in the metric one \cite{0902.4465,0903.0355}. This leads to an interesting possibility to connect inflationary observables with low-energy collider experiments \cite{Shaposhnikov:2020fdv}. As for the (classical) hierarchy problem, \ie the question of why the Higgs mass is so much smaller than the Planck mass, the original proposal towards it is as follows \cite{Shaposhnikov:2018xkv}. One considers a scenario in which the Standard Model is classically scale-invariant, \ie the Higgs mass is set to zero. Furthermore, no degrees of freedom are assumed to exist anywhere above the Electroweak scale. In this situation, it was argued that the Higgs mass can arise due to a non-perturbative gravitational effect. While this was first proposed and studied within the metric formulation of gravity \cite{Shaposhnikov:2018xkv, Shaposhnikov:2018jag, Shkerin:2019mmu}, it turns out that the non-perturbative mechanism of generating the Electroweak scale out of the Planck scale can be implemented more elegantly and naturally in the Palatini version of the theory \cite{Shaposhnikov:2020geh, Karananas:2020qkp}. These observations provide a strong motivation to pursue the study of EC gravity which, as said, is a natural generalization of Palatini gravity in the presence of fermions.

Once fermions are taken into account, the equivalent metric theory, which is obtained by solving for torsion, contains two types of additional higher-dimensional operators. The one represents an interaction between vector and axial fermion currents, and the other one provides a derivative coupling of the Higgs field to fermions. In \cite{Shaposhnikov:2020aen} we show that the fermion current interaction terms can be responsible for production of singlet fermion dark matter in the Early Universe. In this paper, our goal is to investigate whether the terms with fermions affect inflation.

The paper is organized as follows. In section \ref{sec:theory}, we introduce the theory and discuss inflation in general. In the following two sections, we study inflation in two regimes which can be treated analytically. The first one corresponds to switching off the Holst term, $1/\bar{\gamma} = \xi_\gamma =0$. Then, the inflation is parametrized by $\xi$ and $\xi_\eta$ and we call it ``Nieh-Yan inflation''. It is studied in section \ref{sec:NiehYan}. Subsequently, we consider the case in which there is no Nieh-Yan term and no coupling of the Higgs field to the Holst term, $\xi_\gamma=\xi_\eta=0$. 
This leads to the inflation characterized by $\xi$ and $\bar{\gamma}$, hence we adopt the name ``Holst inflation''. It is studied in section \ref{sec:Holst}.
In section \ref{sec:gen}, we perform the full parameter space scan. Section \ref{sec:fermions} contains a discussion of the perturbative cutoff scale in the different models and of the relevance of the scalar-fermion and fermion-fermion interaction terms for inflation. Finally, section \ref{sec:conclusion} is dedicated to discussion and conclusion.

\textbf{Conventions}. We work in natural units $\hbar=c=1$ and use the metric signature $(-1,+1,+1,+1)$. The antisymmetric tensor is defined by $\epsilon_{0123}=1$. The matrix $\gamma^5$ is defined as $\gamma^5=\frac{1}{i}\gamma^0\gamma^1\gamma^2\gamma^3$.

\section{The theory}
\label{sec:theory}

The gravitational part of our theory is displayed in \Eq \eqref{S_grav}. As explained in introduction, the first line is the Einstein-Hilbert action with the Higgs field non-minimally coupled to it; the second line represents the Holst term \cite{Hojman:1980kv, Nelson:1980ph, Castellani:1991et, Holst:1995pc} with the coefficient $\bar{\gamma}$ called the Barbero-Immirzi parameter \cite{Immirzi:1996dr,Immirzi:1996di}, again coupled non-minimally to the Higgs field; and the third line represents the Nieh-Yan invariant (\ref{Nieh-Yan}) \cite{Nieh:1981ww} multiplied by $\xi_\eta h^2$. Overall, we have four independent couplings,
\begin{equation}\label{params}
\xi \;, ~~ \xi_\gamma \;, ~~ \xi_\eta ~~ \rm{and} ~~ \bar{\gamma} \;.
\end{equation}
It will become evident that the theory is healthy provided that all these couplings are real and non-negative. However, more general parameter choices are possible as well for certain regions of field values. 

Next, consider the action of the scalar field $h$:
\begin{equation}
S_h = \int \diff^4x\sqrt{-g}\:\left( -\frac{1}{2}(\partial_\mu h)^2-U(h) \right) \;, ~~~ U(h)=\frac{\lambda}{4}(h^2-v^2)^2 \;.
\end{equation}
Here $\lambda$ is identified with the Higgs quartic coupling and $v$ corresponds to its vacuum expectation value. We will neglect the latter throughout the paper, since it is not relevant during inflation. Finally, bearing in mind the coupling to the full Standard Model, we shall also consider the effect due to fermions. We choose the fermion kinetic term as follows:
\begin{equation}
S_f= \frac{i}{2}\int \diff^4x\sqrt{-g}\left( \bar{\Psi}(1-i\alpha -i\beta\gamma^5)\gamma^\mu D_\mu\Psi - \overline{D_\mu\Psi}(1+i\alpha+i\beta\gamma^5)\gamma^\mu\Psi\right) \;.
\end{equation}
Here we denoted $\Psi=(\phi_\alpha,\chi^{\alpha'})$ and the connection field is contained in the covariant derivative $D$. The real parameters $\alpha$ and $\beta$ are coefficients of ``non-minimal'' kinetic terms. In the absence of torsion, they sum up to a total derivative, but in the torsionful case they contribute to the dynamics of the theory \cite{hep-th/0507253,Alexandrov:2008iy,1104.2432, 1212.0585}. The couplings  $\alpha$ and $\beta$ can be different for different fermion generations.

We would like to bring the theory to the form convenient for the analysis of inflation. To this end, we first get rid of the non-minimal coupling in the Einstein-Hilbert term. This is achieved by the Weyl transformation of the metric field: 
\begin{equation}\label{WeylTransform}
g_{\mu\nu}\to \Omega^2 g_{\mu\nu} \;, ~~~ \Omega^2=1+\frac{\xi h^2}{M_P^2} \;.
\end{equation}
Second, the resulting theory can be explicitly solved for the torsionful part of the connection. By plugging this solution back in the action, one arrives at the effective metric theory. Then the couplings $ \xi, \xi_\gamma, \xi_\eta, \bar{\gamma}$ as well as $\alpha, \beta$ manifest themselves through higher-dimensional operators. Here we quote the result and refer to \cite{Shaposhnikov:2020frq} for its derivation. The total effective action is
\begin{equation}\label{S_eff}
S_{\rm tot}^{\rm eff}=S_{\rm metric}+S_{hVA} \;,
\end{equation}
where $S_{\rm metric}$ is the part of the action that one would have also obtained in metric gravity:
\begin{equation}\label{S_metric}
\begin{split}
S_{\rm metric}= 
&\int \diff ^4x\sqrt{-g}\left\lbrace \frac{M_P^2}{2}\mathring{R}+\frac{i}{2}\overline{\Psi}\gamma^\mu\mathring{D}_\mu\Psi-\frac{i}{2}\overline{\mathring{D}_\mu\Psi}\gamma^\mu\Psi \right\rbrace \\
& - \int d^4x\sqrt{-g} \left\lbrace \frac{1}{2\Omega^2}(\partial_\mu h)^2+\frac{U}{\Omega^4} \right\rbrace\;.
\end{split}
\end{equation}
The superscript in $\mathring{R}$ and $\mathring{D}_\mu$ indicates that the corresponding quantities are evaluated with $T^\mu_{\;\;\nu\rho}=0$.

The additional higher-order terms are given by
\begin{equation}\label{S_hVA}
\begin{split}
S_{hVA} & = -\int \diff^4x\sqrt{-g}\frac{3M_P^2}{4(\gamma^2+1)}\left( \frac{\partial_\mu\bar{\eta}}{\Omega^2}+\partial_\mu\gamma \right)^2\\
& + \int \diff^4x\sqrt{-g}\frac{3\alpha}{4}\left( \frac{\partial_\mu\Omega^2}{\Omega^2}+\frac{\gamma}{\gamma^2+1}\left( \frac{\partial_\mu\bar{\eta}}{\Omega^2}+\partial_\mu\gamma \right) \right) V^\mu \\
& + \int \diff^4x\sqrt{-g}\frac{3}{4}\left( \beta\frac{\partial_\mu\Omega^2}{\Omega^2}+\frac{1+\gamma\beta}{\gamma^2+1}\left( \frac{\partial_\mu\bar{\eta}}{\Omega^2}+\partial_\mu\gamma \right) \right) A^\mu \\
& - \int \diff^4x\sqrt{-g}\frac{3}{16M_P^2(\gamma^2+1)}\left((1+2\beta\gamma-\beta^2)A_\mu^2+2\alpha(\gamma-\beta)A_\mu V_\mu-\alpha^2 V_\mu^2\right) \;,
\end{split}
\end{equation}
where
\begin{equation}
V_\mu=\bar{\Psi}\gamma_\mu\Psi \;, ~~~ A_\mu =\bar{\Psi}\gamma^5\gamma_\mu\Psi
\end{equation}
are vector and axial fermion currents, correspondingly, and we defined
\begin{equation}\label{GammaEta}
\gamma=\frac{1}{\bar{\gamma}\Omega^2}\left(1+\frac{\xi_\gamma h^2}{M_P^2}\right) \;, ~~~ \bar{\eta}=\frac{\xi_\eta h^2}{M_P^2} \;.
\end{equation}

All features of the theory relevant for inflation are now encoded in the terms containing derivatives of $h$. Combining the first line of \Eq (\ref{S_hVA}) with the Higgs field kinetic term from (\ref{S_metric}), we find that the total kinetic term is
\begin{equation}
-\frac{1}{2}K(h) (\partial_\mu h)^2 \;,
\end{equation}
where
\begin{equation} \label{modifiedK}
K(h) = \frac{1}{\Omega^2} \left(1 +\frac{ 6h^2 \left(\frac{\xi_\gamma -\xi}{\bar{\gamma}} + \xi_\eta \Omega^2\right)^2}{M_P^2\Omega^2 \left(\Omega^4 + \frac{1}{\bar{\gamma}^2}\left(1 + \frac{\xi_\gamma h^2}{M_P^2}\right)^2\right)}\right) \,,
\end{equation}
and we made use of \Eqs (\ref{GammaEta}). Clearly, the kinetic term is positive-definite as soon as the couplings (\ref{params}) are all non-negative.
Now we introduce the canonical field variable $\chi$ defined by
\begin{equation} \label{chi}
\chi = \limitint_0^h dh'\:\sqrt{K(h')} \;.
\end{equation}
Then the action for the scalar field becomes simply
\begin{equation}\label{PotChi}
S_{h} = \int \diff^4x\sqrt{-g}\:\left( -\frac{1}{2}(\partial_\mu\chi)^2-U(\chi) \right) \;, ~~~ U(\chi)=\frac{\lambda h(\chi)^4}{4\Omega(\chi)^4} \;.
\end{equation}
Since $\chi$ is a monotonic function of $h$, we have $U'(\chi)>0$ for $\chi>0$. Thus, $U(\chi)$ is a valid inflationary potential which develops a flat direction at large $\chi$ or $h$. 

From the above we see similarities with the known Higgs inflation models \cite{Bezrukov:2007ep, Bauer:2008zj}. In fact, both the metric and Palatini versions of Higgs inflation are reproduced at certain values of the couplings (\ref{params}). To see this, consider first the limit of vanishing Holst term, $\bar{\gamma}\to\infty$, $\xi_\gamma=0$. Then \Eq (\ref{modifiedK}) gives
\begin{equation}\label{NY_K}
\left. K(h)\right\vert_{\bar{\gamma}\to\infty,~\xi_\gamma=0}=\frac{1}{\Omega^2}+\frac{6\xi_\eta^2h^2}{\Omega^4M_P^2} \;.
\end{equation}
Taking further the limit $\xi_\eta=0$, we recover the Palatini formulation of the Higgs inflation theory. On the other hand, if $\xi_\eta=\xi$, the metric formulation is reproduced. Thus, by varying continuously the Higgs coupling to the Nieh-Yan term, one can deform one model into another. Next, we consider the limit $\bar{\gamma}=\xi_\gamma=\xi_\eta=0$. One can show that this limit implies vanishing torsion \cite{ hep-th/0507253,Shaposhnikov:2020frq}. This can be seen directly by inspecting \Eq (\ref{modifiedK}):
\begin{equation}\label{Metric_K}
\left. K(h)\right\vert_{\bar{\gamma}=\xi_\gamma=\xi_\eta=0}=\frac{1}{\Omega^2}+\frac{6\xi^2h^2}{\Omega^4M_P^2} \;.
\end{equation}
Thus, we recover metric Higgs inflation.

\section{Nieh-Yan inflation}
\label{sec:NiehYan}

The Nieh-Yan inflation is described by the kinetic term (\ref{NY_K}) obtained from the general expression (\ref{modifiedK}) in the limit of vanishing Holst term, $\bar{\gamma}\to\infty$, $\xi_\gamma=0$. For the analytic study, we work under the assumption
\begin{equation}\label{NY_cond}
 h\gg \frac{M_P}{\sqrt{\xi}} \;.
\end{equation}
It will turn out that this inequality holds during inflation in all relevant parts of the parameter space. Using \Eq (\ref{NY_cond}), one can perform the integral in \eqref{chi} explicitly and obtain the relation between $h$ and the canonical field $\chi$:
\begin{equation} \label{NY_chi}
	h = \frac{M_P}{\sqrt{\xi}} \exp\left\{\frac{\sqrt{\xi}\chi}{M_P \sqrt{1 + \frac{6 \xi_\eta^2}{\xi}}}\right\}\,.
\end{equation}
Plugging this into the potential, we obtain its form during inflation:
\begin{equation}\label{NY_PotInfl}
U(\chi)=\frac{\lambda M_P^4}{4\xi^2}\left( 1+\exp\left\{-\frac{2\xi}{\sqrt{\xi+6\xi_\eta^2}}\frac{\chi}{M_P}\right\} \right)^{-2} \;.
\end{equation}
Using \Eq (\ref{NY_PotInfl}), it is straightforward to compute the slow-roll parameters; they are given by
\begin{align}
   & \epsilon=\frac{M_P^2}{2}\left(\frac{U'}{U}\right)^2=\frac{8M_P^4}{\xi h^4\left(1+\frac{6\xi_\eta^2}{\xi}\right)} \;, \label{NY_eps}\\
   & \eta=M_P^2\frac{U''}{U}=-\frac{8M_P^2}{h^2\left(1+\frac{6\xi_\eta^2}{\xi}\right)} \;, \label{NY_eta}
\end{align}
where prime denotes differentiation with respect to $\chi$. The value of $h$ at which either $\epsilon$ or $|\eta|$ becomes of the order of $1$ marks the end of inflation. According to \Eqs (\ref{NY_eps}) and (\ref{NY_eta}), the parameter $|\eta|$ grows faster than $\epsilon$ as $h$ decreases. Thus, the field value at the end of inflation is
\begin{equation}\label{NY_hend}
    h_{\rm end}=\frac{\sqrt{8}M_P}{\sqrt{1+\frac{6\xi_\eta^2}{\xi}}} \;.
\end{equation}
Next, we compute the number of e-foldings as a function of the field value $h$:
\begin{equation}\label{N_gen}
    N_\star=\frac{1}{M_P^2}\limitint_{h_{\rm end}}^{h_\star} dh\sqrt{K}\frac{U}{U'}=\frac{1}{8M_P^2}\left(1+\frac{6\xi_\eta^2}{\xi}\right)(h_\star^2-h_{\rm end}^2) \;.
\end{equation}
To the leading order in $1/N_\star$, this leads to
\begin{equation}\label{NY_h}
    h_\star=M_P{\sqrt{\frac{8N_\star}{1+\frac{6\xi_\eta^2}{\xi}}}} \;.
\end{equation}
We substitute this into \Eqs (\ref{NY_eps}) and (\ref{NY_eta}) to find the spectral index and the tensor-to-scalar ratio:
\begin{align}
    & n_s=1-6\epsilon+2\eta=1-\frac{2}{N_\star}-\frac{3(\xi+6\xi_\eta^2)}{4N_\star^2\xi^2} \label{NY_ns}\;, \\
    & r=16\epsilon=\frac{2(\xi+6\xi_\eta^2)}{N_\star^2\xi^2} \;. \label{NY_r}
\end{align}

In order to compute numerical values, we first implement the constraint due to the CMB-normalization \cite{Akrami:2018odb},  
\begin{equation}\label{CMB_norm}
U/\epsilon = 5.0 \cdot 10^{-7} M_P^4 \;,
\end{equation}
which refers to the scale $k_\star=0.05$ Mpc$^{-1}$. To the leading order in $1/N_\star$, this gives
\begin{equation} \label{normalization}
\frac{2 \lambda N_\star^2}{\xi +  6\xi_\eta^2} = 5.0 \cdot 10^{-7} \;.
\end{equation}
As a final ingredient, we need the number $N_\star$ of e-foldings, at which CMB perturbations are generated. This, in turn, requires knowledge of post-inflationary dynamics of the Universe including the duration and temperature of preheating.\footnote
{In Palatini Higgs inflation, preheating occurs instantaneously \cite{Rubio:2019ypq}. It has been argued that the same is true in metric Higgs inflation, although this conclusion is not final since the temperature of preheating lies above the perturbative cutoff in the metric scenario \cite{Ema:2016dny, DeCross:2016cbs}.} 
 Studying preheating in the theory of Nieh-Yan inflation goes beyond the scope of the present paper. Instead, we take
\begin{equation}\label{N}
N_\star=55 
\end{equation}
as a simple estimate. Our results will be only mildly sensitive to the precise value of $N_\star$. Moreover, we note that in both metric and Palatini Higgs inflation $N_\star$ lies fairly close to the value (\ref{N}); see, e.g., \cite{Shaposhnikov:2020fdv}. 

Now we can analyze different regimes of Nieh-Yan inflation. From \Eqs (\ref{NY_ns}), (\ref{NY_r}) and \eqref{normalization} we see that three regimes and one special case are possible:

$\bullet$ $\xi_\eta\lesssim\sqrt{\xi}$. In this limit, $\xi_\eta$ can be neglected and we reproduce Palatini Higgs inflation. In agreement with the known results \cite{Bauer:2008zj}, \Eq \eqref{normalization} leads to $\xi\approx 1\cdot 10^{10}\lambda$ and the inflationary indices become $n_s=1-2/N_\star$ and $r=2/(\xi N_\star^2)$. 

$\bullet$ $\sqrt{\xi}\lesssim\xi_\eta\lesssim \xi$. In this regime, it follows from \Eq \eqref{normalization} that the value of $\xi_\eta$ is fixed by the amplitude of CMB perturbations: 
\begin{equation}\label{XiEtaNY}
\xi_\eta\approx 4\cdot 10^4\sqrt{\lambda} \;.
\end{equation}
Thus, $\xi_\eta$ has the numerical value that $\xi$ assumes in metric Higgs inflation.
For the spectral indices we have
\begin{equation}\label{NY_ns_r}
    n_s=1-\frac{2}{N_\star} \;, ~~~ r=\frac{12\xi_\eta^2}{N_\star^2\xi^2} \;.
\end{equation}
We see that $n_s$ is still independent of both $\xi$ and $\xi_\eta$, like in metric and Palatini Higgs inflation. In contrast, $r$ depends on the parameter $\xi$. The latter can be varied freely in the range $\xi_\eta\lesssim\xi\lesssim\xi_\eta^2$. If the upper bound is violated, $\xi$ is no longer independent of the CMB normalization. The lower bound is due to the observational constraint on the tensor-to-scalar ratio \cite{Akrami:2018odb,Tanabashi:2018oca},
\begin{equation}\label{R}
r<0.06 \;.
\end{equation}

$\bullet$ $\xi_\eta=\xi$. At this special value of the Nieh-Yan coupling we recover the original metric Higgs inflation. This can be seen directly from the structure of the kinetic term (\ref{NY_K}). In this limit, we obtain $n_s=1-2/N_\star$ and $r=12/N_\star^2$, in agreement with previous studies \cite{Bezrukov:2007ep}.

$\bullet$ $\xi\lesssim\xi_\eta$. In this regime, the approximation \eqref{NY_cond} no longer holds so that our analytic study cannot be applied. However, extrapolating the expression \Eq \eqref{NY_ns_r} for the tensor-to-scalar ratio, one can expect that the latter is too big to be compatible with the observational bound.

\begin{figure}[t]
\begin{center}
\begin{minipage}[h]{0.49\linewidth}
\center{\includegraphics[width=\textwidth]{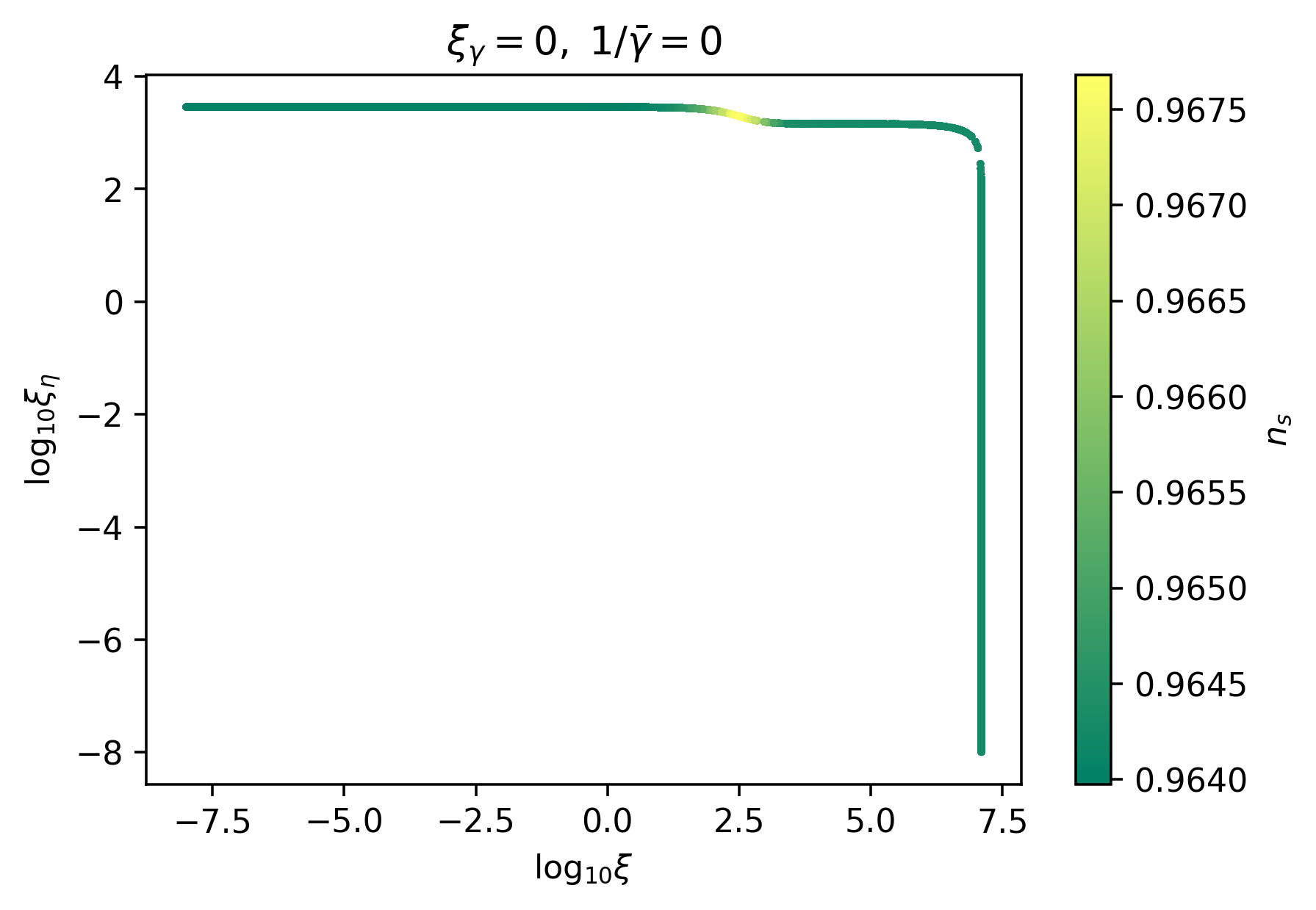} \\(a)}
\end{minipage}
\begin{minipage}[h]{0.49\linewidth}
\center{\includegraphics[width=\textwidth]{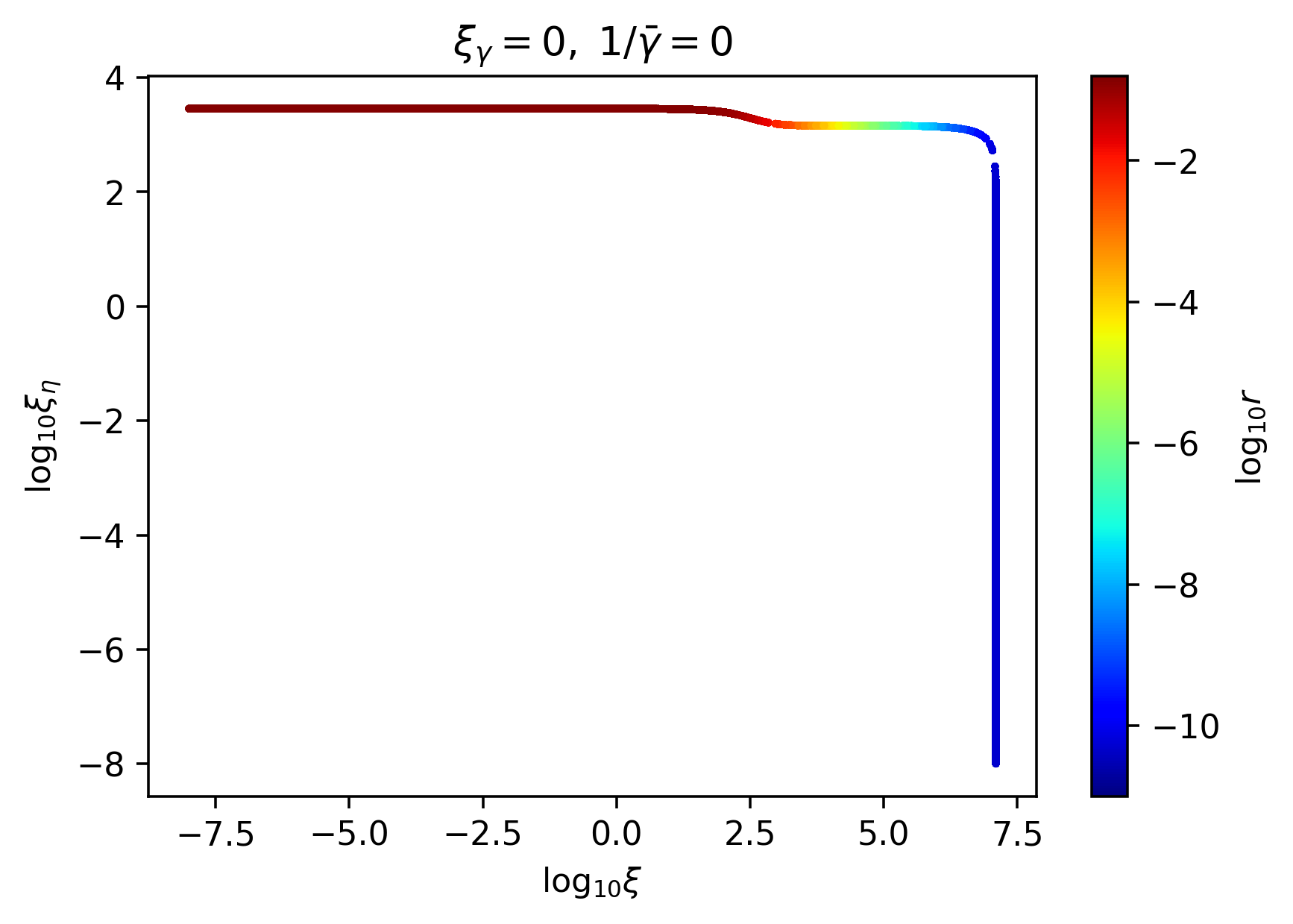} \\(b)}
\end{minipage}
\caption{Spectral tilt (a) and tensor-to-scalar ratio (b) in Nieh-Yan inflation. We take $N_\star=55$ and $\lambda=10^{-3}$. The regions of Palatini (the right vertical segment) and metric (the ``ankle'' at which $n_s$ and $r$ vary considerably) Higgs inflation are clearly distinguishable. The transition between the two regions is smooth and stays within the observational bounds. The left horizontal segment  has $r> 0.1$ and is not compatible with observations. }
\label{fig:NY}
\end{center}
\end{figure}

\begin{figure}[h]
    \centering
    \includegraphics[width=0.45\textwidth]{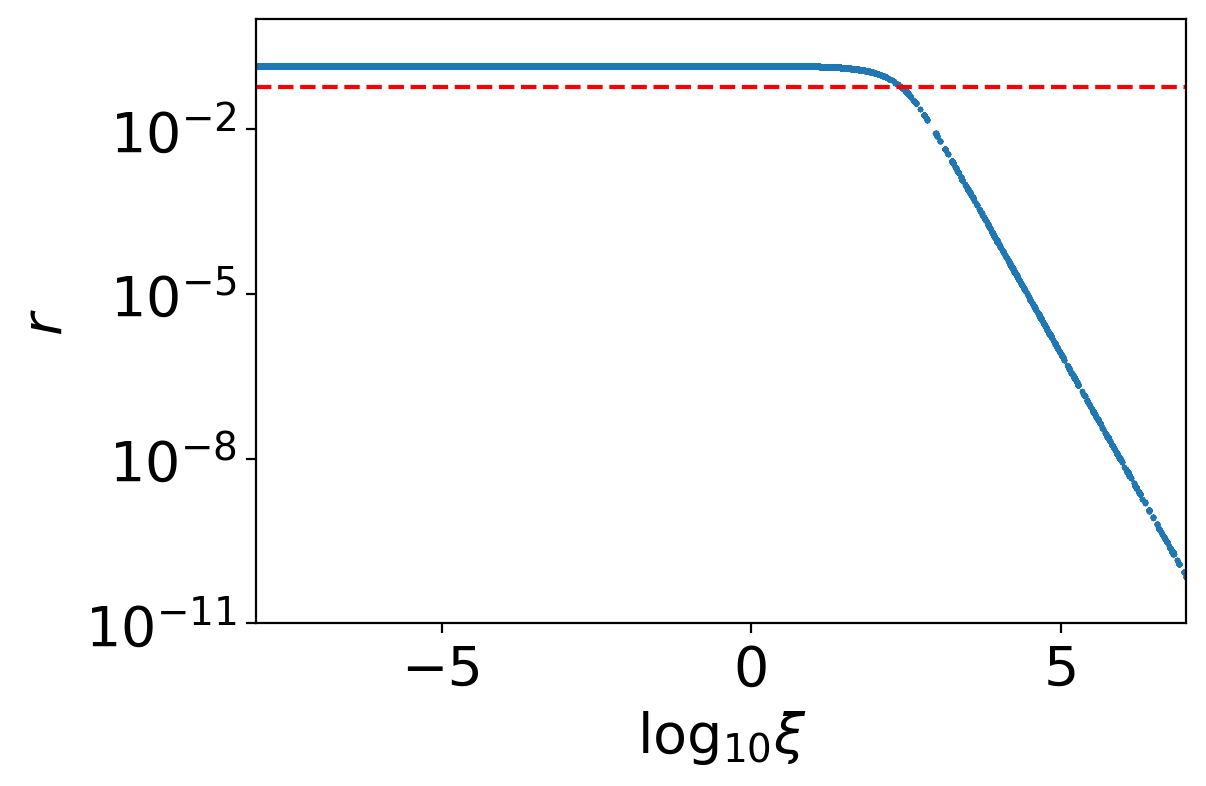}
    \caption{Tensor-to-scalar ratio $r$ in Nieh-Yan inflation as a function of $\xi$. The range of $\xi$ corresponds to the horizontal branch shown in figure~\ref{fig:NY} (b). The red dashed line shows the limit $r=0.06$. The values of $r$ above this line are excluded by observations. The horizontal region of the blue line corresponds to the value $r=0.144$.}
    \label{fig:NYr}
\end{figure}

In figure \ref{fig:NY}, we explore numerically the parameter space of the Nieh-Yan inflation. In all subsequent numerical estimations we take $\lambda$ equal a typical value of the Higgs quartic coupling during inflation:\footnote{The Higgs self-coupling constant at inflationary scale $\mu_{\rm inf}$ is much smaller than its value at low energies because of renormalisation group running. Within the experimental uncertainties in the top quark Yukawa coupling (for a discussion see, e.g., \cite{Bezrukov:2014ina}), the  actual value of $\lambda$ at $\mu_{\rm inf}$ varies roughly between  -0.01 and +0.01. The choice we made in \Eq\eqref{Lambda} is fully compatible with the recent measurements  of the top quark pole mass at ATLAS \cite{Sirunyan:2019zvx} and CMS \cite{Aad:2019mkw}, $m_{t}^{\text{pole}}=170.5\pm0.8\,\text{GeV}$  and $m_{t}^{\text{pole}}=171.1\pm1.2\,\text{GeV}$ respectively. In addition, it corresponds to the preferred value we found in \cite{Shaposhnikov:2020fdv} while analysing quantum effects in Palatini Higgs inflation. For a discussion of the Higgs inflation with negative values of $\lambda$ see \cite{Bezrukov:2014ipa}.}
\begin{equation}\label{Lambda}
\lambda=10^{-3} \;.
\end{equation}
We use the ensemble Markov chain Monte Carlo (MCMC) method to scan over positive values of the couplings $\xi$ and $\xi_\eta$ and select those yielding \Eq (\ref{N}). For details of our procedure, see appendix \ref{sec:numerical_procedure}. Our numerical analysis confirms the results of analytic study made above. We observe the regions of Palatini and metric Higgs inflation as well as a smooth interpolation between them as $\xi_\eta$ grows from $\sqrt{\xi}$ to $\xi$. We see also that at even larger values of $\xi_\eta$ the tensor-to-scalar ratio $r$ continues to grow and quickly becomes incompatible with observations. It is interesting to note, however, that the observational bound (\ref{R}) is saturated at ${\xi\simeq 261, \;\xi_\eta\simeq 2060}$, not far from the metric Higgs inflation limit. We show how $r$ depends on $\xi$ in figure~\ref{fig:NYr}. 

\section{Holst inflation}
\label{sec:Holst}

Let us now study inflation driven by the Holst term without non-minimal coupling to the Higgs field. Correspondingly, we set $\xi_\gamma=\xi_\eta=0$. As is evident from \Eq \eqref{modifiedK}, the resulting kinetic term is
\begin{equation}\label{Holst_K}
\left. K(h)\right\vert_{ \xi_\gamma=\xi_\eta=0 }=\frac{1}{\Omega^2}\left( 1+\frac{6\xi^2h^2}{M_P^2\bar{\gamma}^2\Omega^2(\Omega^4+\frac{1}{\bar{\gamma}^2})} \right) \;.
\end{equation}
Below we investigate it analytically in different parameter ranges.

$\bullet$ $1/\bar{\gamma} \gtrsim N_\star$. As discussed in section \ref{sec:theory}, the metric formulation of the theory is recovered in the limit $\bar{\gamma}\to 0$ (see \Eq (\ref{Holst_K})). In order to make a more quantitative statement, we note that the kinetic term of the metric formulation is reproduced as long as
\begin{equation}\label{MetricCond}
\bar{\gamma}^2\Omega^4\ll 1 \;.
\end{equation}
In the metric formulation, $h\sim \sqrt{N} M_P/\sqrt{\xi}$, where $N$ is the number of e-foldings before the end of inflation.  Substituting this into \Eq (\ref{MetricCond}), we find that the number of e-foldings is bounded as $N\lesssim N_{\rm metric}$, where
\begin{equation}\label{borderGamma}
N_{\text{metric}} = \frac{1}{\bar{\gamma}}\,.
\end{equation}
In other words, the Barbero-Immirzi parameter $1/\bar{\gamma}$ sets the maximal duration of the metric phase of inflation that the potential can support. Fixing the number of e-foldings corresponding to the CMB epoch according to \Eq (\ref{N}), we conclude that the CMB perturbations are produced during the metric phase of inflation as long as $1/\bar{\gamma} \gtrsim N_\star$.

$\bullet$ $1/\sqrt{\xi} \lesssim 1/\bar{\gamma} \lesssim N_\star$. We know from the discussion in section \ref{sec:theory} that the Palatini formulation of the theory is obtained in the limit $\bar{\gamma} \rightarrow \infty$. Again, we want to make a more precise statement. To this end, we note that one can neglect the second term in \Eq \eqref{Holst_K} as long as 
\begin{equation}\label{PalatiniCond}
\frac{\xi^2h^2}{\bar{\gamma}^2\Omega^6M_P^2} \ll1 \;,~~~ \frac{1}{\bar{\gamma}}\lesssim \Omega^2 \;.
\end{equation}
Using that $h\gtrsim M_P$ in Palatini Higgs inflation, from \Eqs (\ref{PalatiniCond}) we conclude that the Palatini phase of inflation takes place for $1/\bar{\gamma} \lesssim \sqrt{\xi}$. Furthermore, taking into account the consideration of the metric phase in the previous point, we conclude that for $1 \lesssim 1/\bar{\gamma} \lesssim \sqrt{\xi}$, the potential supports the regions of both metric and Palatini Higgs inflation, with the latter taking place at higher values of $h$ (or $\chi$). 

Thus, as long as $1/\bar{\gamma} \gtrsim N_\star$, the inflaton behaves as in the metric scenario during the last $N_\star$ e-foldings, making the model observationally equivalent to metric Higgs inflation. In contrast, if $1/\bar{\gamma} \lesssim N_\star$, the CMB perturbations are generated during a period of Palatini Higgs inflation. It should be noted, however, that in the latter case the dynamics of the inflaton at $N<N_\star$ is different from the Palatini scenario, and there is no full equivalence between the two models. 

$\bullet$ $1/\bar{\gamma} \lesssim 1/\sqrt{\xi}$. It is natural to ask under what conditions the Holst inflation model is fully equivalent to the Palatini model. The answer is that the scalar field kinetic term must be close to $1/\Omega^2$ at all $h$. This is because the relation (\ref{chi}) between $h$ and the canonical field $\chi$ is, in principle, sensitive to all values of $h$. Estimating the expression (\ref{Holst_K}) at $h\sim M_P/\sqrt{\xi}$ gives the strongest condition on the Barbero-Immirzi parameter:
\begin{equation}
\frac{1}{\bar{\gamma}}\lesssim \frac{1}{\sqrt{\xi}} \;.
\end{equation}
Only in this case the potential, as a function of $\chi$, corresponds to the one in Palatini Higgs inflation.

\begin{figure}[t]
	\begin{center}
		\begin{minipage}[h]{0.49\linewidth}
			\center{\includegraphics[width=\textwidth]{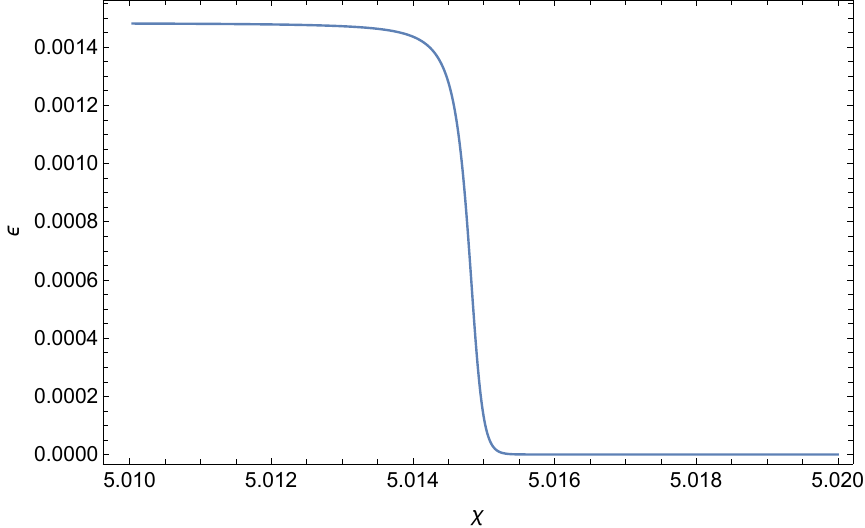} \\(a)}
		\end{minipage}
		\begin{minipage}[h]{0.49\linewidth}
			\center{\includegraphics[width=\textwidth]{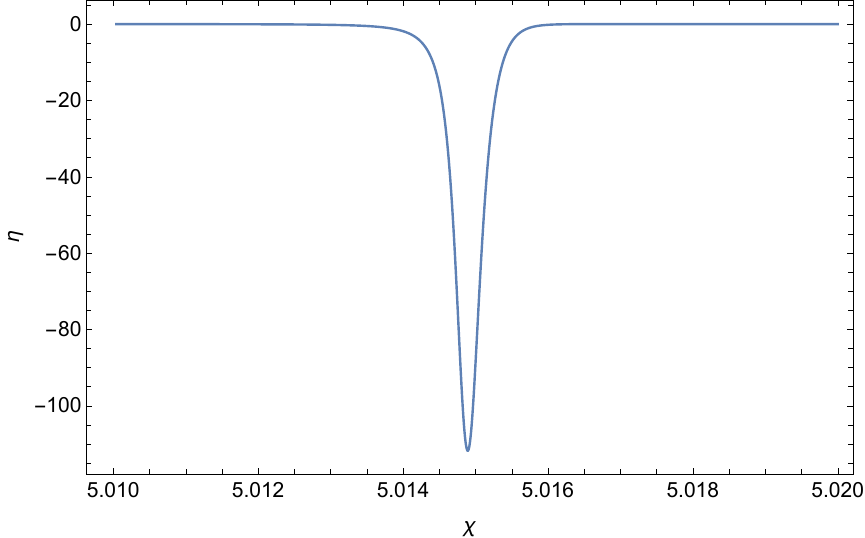} \\(b)}
		\end{minipage}
		\caption{The slow-roll parameters $\epsilon = (U'/U)^2/2$ (a) and 
			$\eta=U''/U$ (b) defined as functions of $\chi$ in the case of Holst inflation for $1/\bar{\gamma} = 30$ and $\lambda=10^{-3}$. All quantities are in units of $M_P$. We see that at $\chi\gtrsim 5.015$ the inflation proceeds as in the Palatini scenario and at $\chi\lesssim 5.015$ -- as in the metric scenario. In the brief transition region, the slow-roll is violated. }
		\label{fig:HolstTransition}
	\end{center}
\end{figure}

It is interesting to look in detail at the transition between the metric and Palatini phases of inflation by integrating numerically \Eq \eqref{chi} and obtaining the potential $U$ as a function of $\chi$. In figure \ref{fig:HolstTransition}, we show the result for an exemplary value of $1/\bar{\gamma}$ in the region $1/\sqrt{\xi} \lesssim 1/\bar{\gamma} \lesssim N_\star$. We observe that the slow-roll parameter $\epsilon$ interpolates monotonically between the Palatini and metric values. As for $\eta$, it develops a spike in the transition region, where $|\eta|$ briefly becomes big. The slow-roll is, therefore, disrupted in this region, hence the values of $1/\bar{\gamma}$ very close to $N_\star$ are not viable.

\begin{figure}[h]
	\begin{center}
		\begin{minipage}[h]{0.49\linewidth}
			\center{\includegraphics[width=\textwidth]{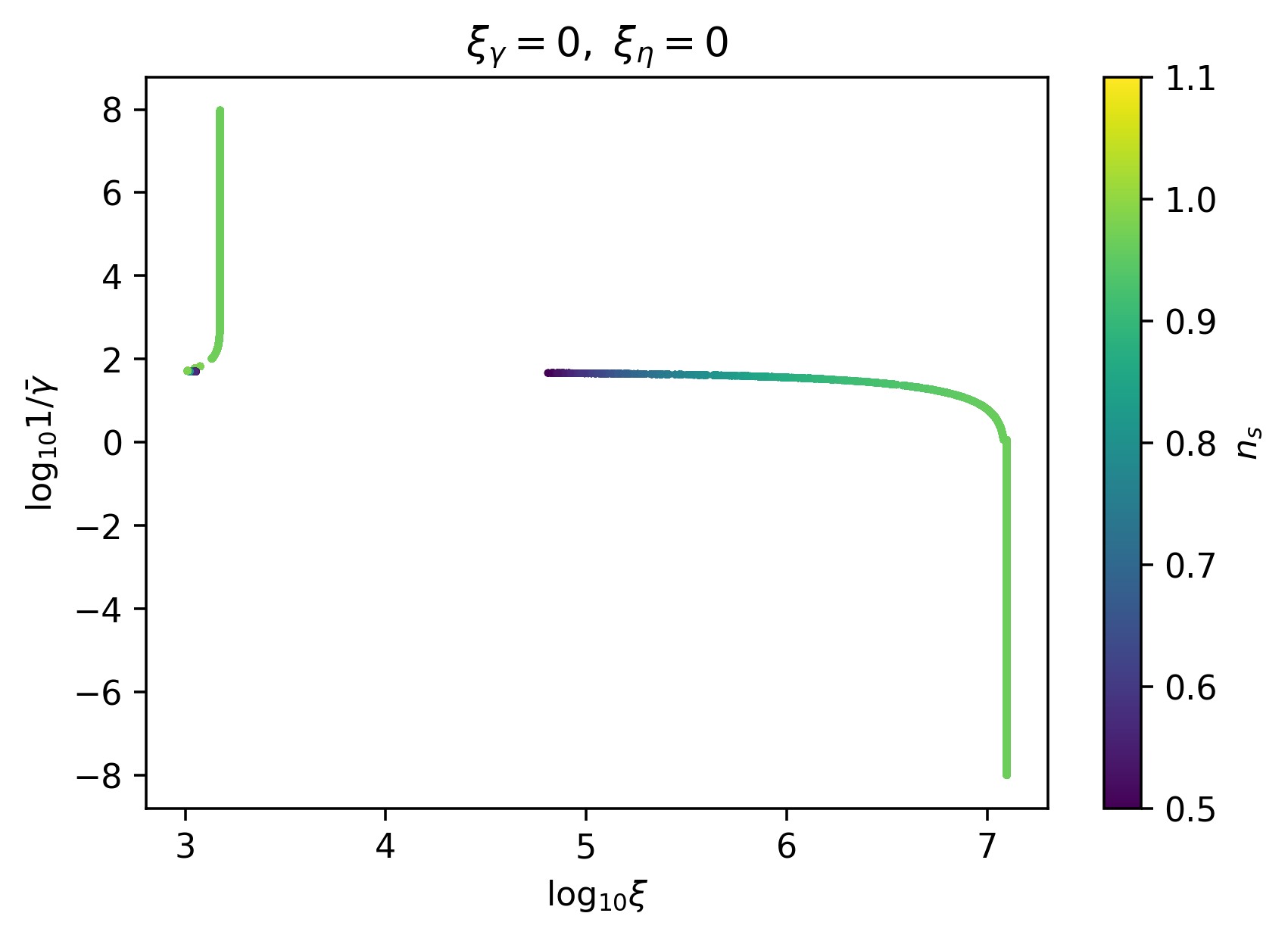} \\(a)}
		\end{minipage}
		\begin{minipage}[h]{0.49\linewidth}
			\center{\includegraphics[width=\textwidth]{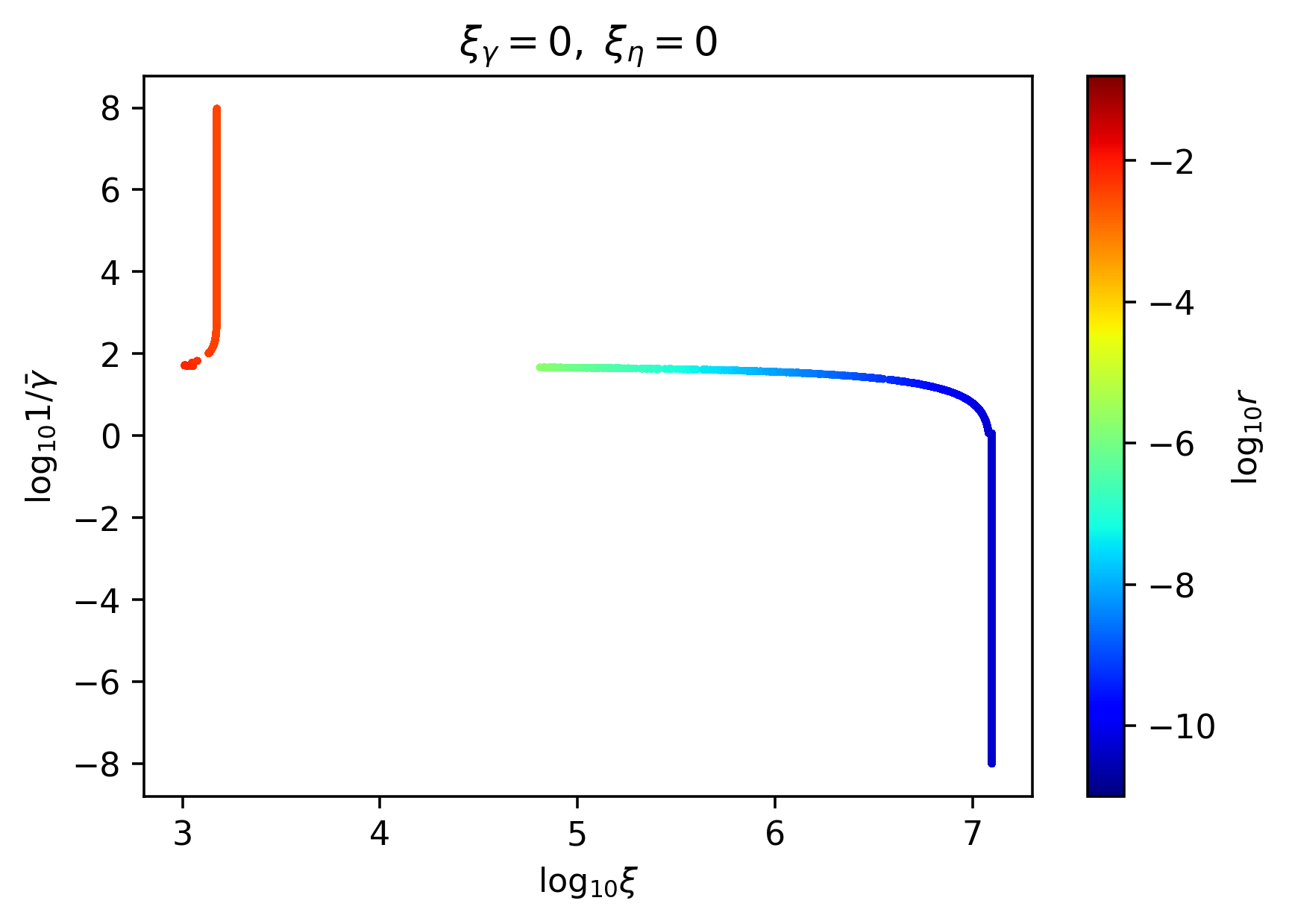} \\(b)}
		\end{minipage}
		\caption{Spectral tilt (a) and tensor-to-scalar ratio (b) in Holst inflation. We take $N_\star=55$ and $\lambda=10^{-3}$. The left-top and the right-bottom vertical segments reproduce the metric and Palatini Higgs inflation, correspondingly. The horizontal segment shows the phenomenologically acceptable part of the connection region. We show only those points in the parameter space which yield $n_s>0.5$. }
		\label{fig:Holst}
	\end{center}
\end{figure}

In figure \ref{fig:Holst}, we present the results of numerical analysis of the full parameter space of Holst inflation. Again, we adopt the MCMC approach and scan over all positive values of $\xi$ and $1/\bar{\gamma}$ requiring that \Eq (\ref{N}) holds;
see appendix \ref{sec:numerical_procedure} for details. We remark that although slow-roll is briefly violated after the generation of CMB perturbations for $1\lesssim 1/\bar{\gamma} \lesssim N_\star$, we continue to use slow-roll equations to compute $N_\star$. This is expected to lead to a small error, which may also explain the discontinuity in \fig \ref{fig:Holst}.  We exclude the points at which the slow-roll conditions are violated during the generation of CMB, by imposing $n_s>0.5$. We observe the two regions corresponding to the metric and Palatini Higgs inflation. The Barbero-Immirzi parameter $1/\bar{\gamma}$ serves to switch between the two limits; otherwise the dependence on it is very mild. We see also that, unlike the Nieh-Yan case, the connection region between the metric and Palatini limits (in which $1/\bar{\gamma} \approx N_\star$) is not phenomenologically viable. These observations confirm the results of the analytic study above.

\section{The general case}
\label{sec:gen}

\begin{figure}[h]
    \begin{center}
        \begin{minipage}[h]{0.49\linewidth}
            \center{\includegraphics[width=\textwidth]{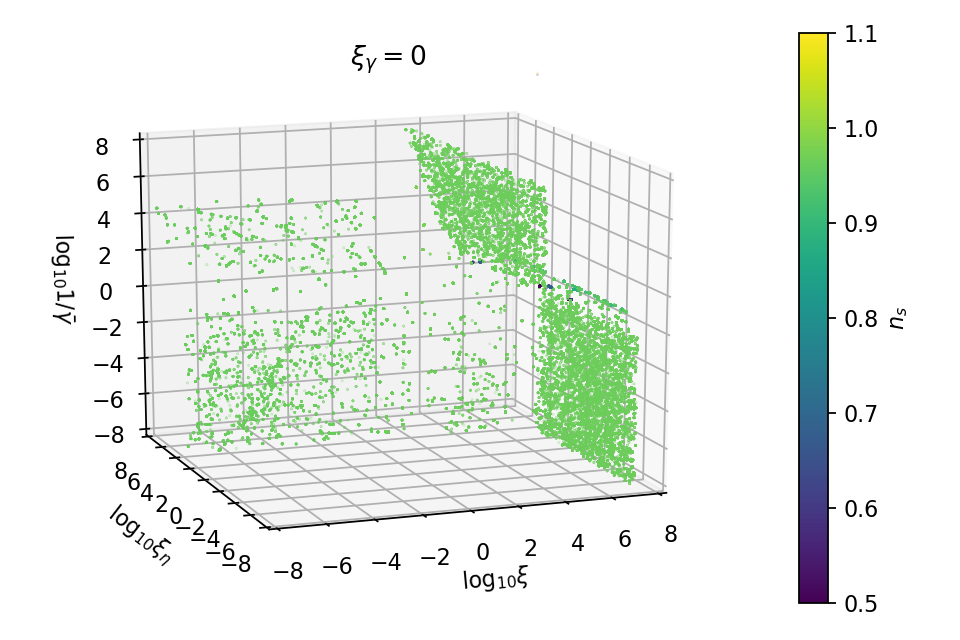} \\(a)}
        \end{minipage}
        \begin{minipage}[h]{0.49\linewidth}
            \center{\includegraphics[width=\textwidth]{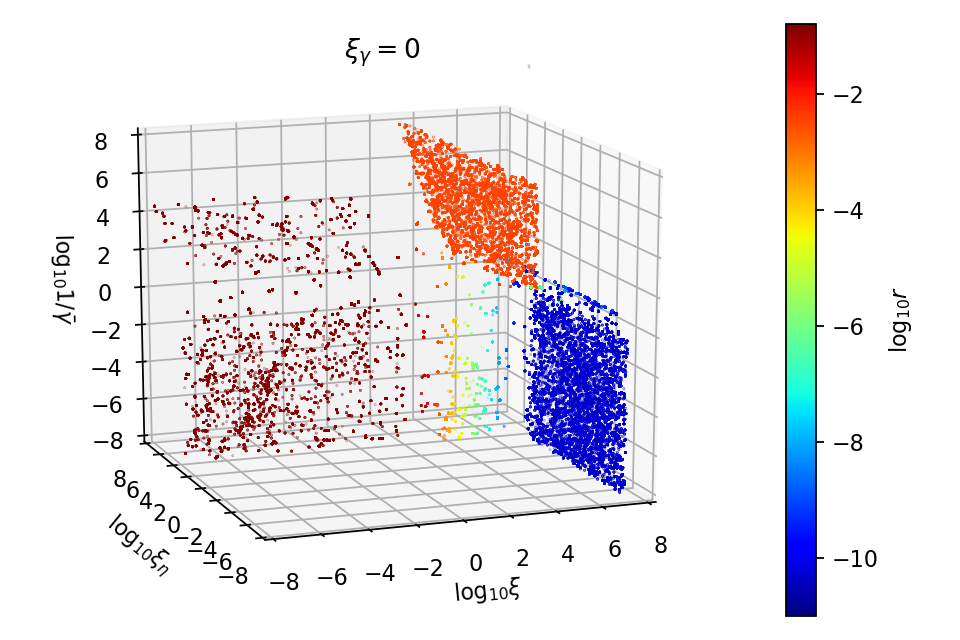} \\(b)}
        \end{minipage}
        \caption{Spectral tilt (a) and tensor-to-scalar ratio (b) in the case $\xi_\gamma = 0$.  One can see that two regions in the right part of the plots reproduce metric and Palatini Higgs inflation. The left region is completely new. Note that due to the large values of the tensor-to-scalar ratio, this region is observationally excluded.}
        \label{fig:Gen1}
    \end{center}
\end{figure}

\begin{figure}[h]
    \begin{center}
        \begin{minipage}[h]{0.49\linewidth}
            \center{\includegraphics[width=\textwidth]{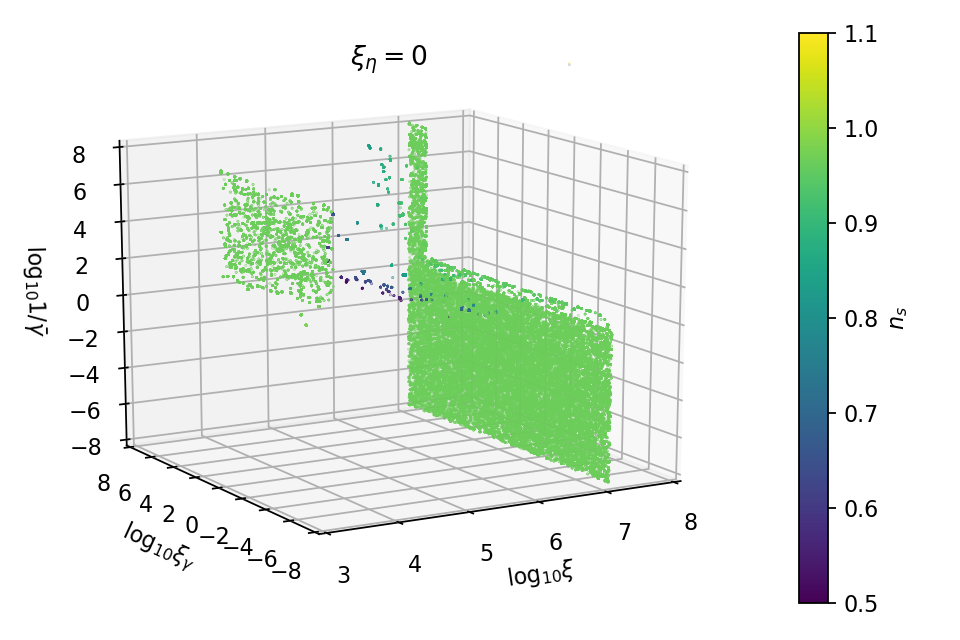} \\(a)}
        \end{minipage}
        \begin{minipage}[h]{0.49\linewidth}
            \center{\includegraphics[width=\textwidth]{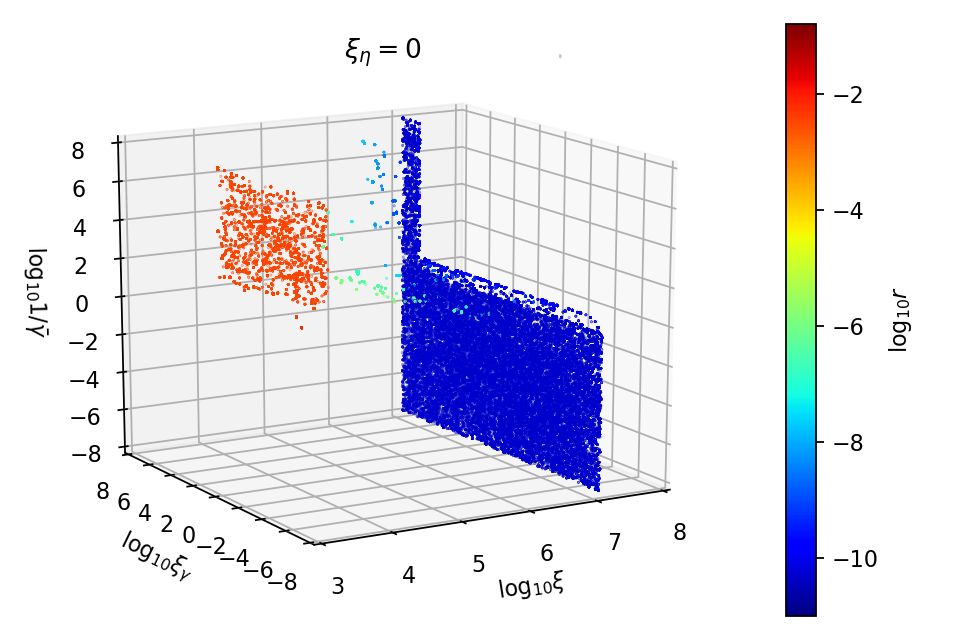} \\(b)}
        \end{minipage}
        \caption{Spectral tilt (a) and tensor-to-scalar ratio (b) in the  case $\xi_\eta = 0$.  The right and left parts of the plots correspond to generalisations of the Palatini and metric Higgs inflation, respectively.}
        \label{fig:Gen2}
    \end{center}
\end{figure}

\begin{figure}[h]
    \begin{center}
        \begin{minipage}[h]{0.49\linewidth}
            \center{\includegraphics[width=\textwidth]{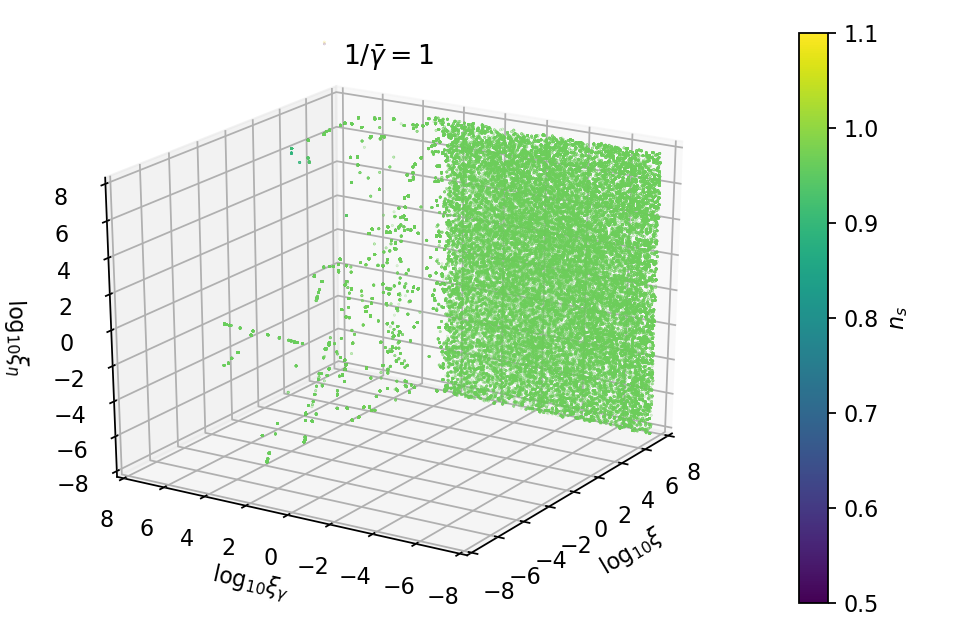} \\(a)}
        \end{minipage}
        \begin{minipage}[h]{0.49\linewidth}
            \center{\includegraphics[width=\textwidth]{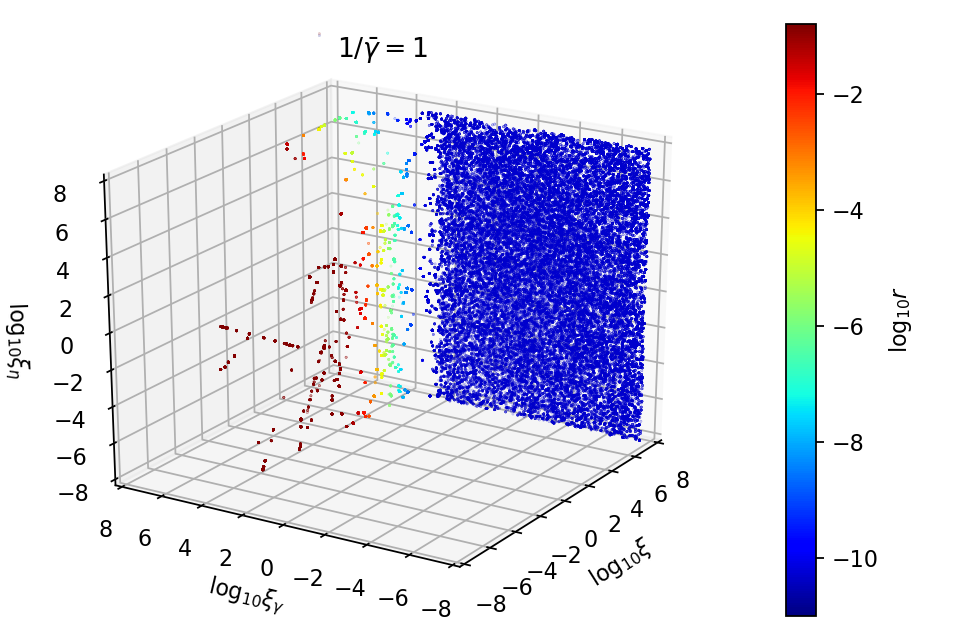} \\(b)}
        \end{minipage}
        \caption{Spectral tilt (a) and tensor-to-scalar ratio (b) in the  case $1/ \bar{\gamma} = 1$. In this case all three non-minimal couplings play a role.}
        \label{fig:Gen3}
    \end{center}
\end{figure}

\begin{figure}[h]
    \begin{center}
        \begin{minipage}[h]{0.49\linewidth}
            \center{\includegraphics[width=\textwidth]{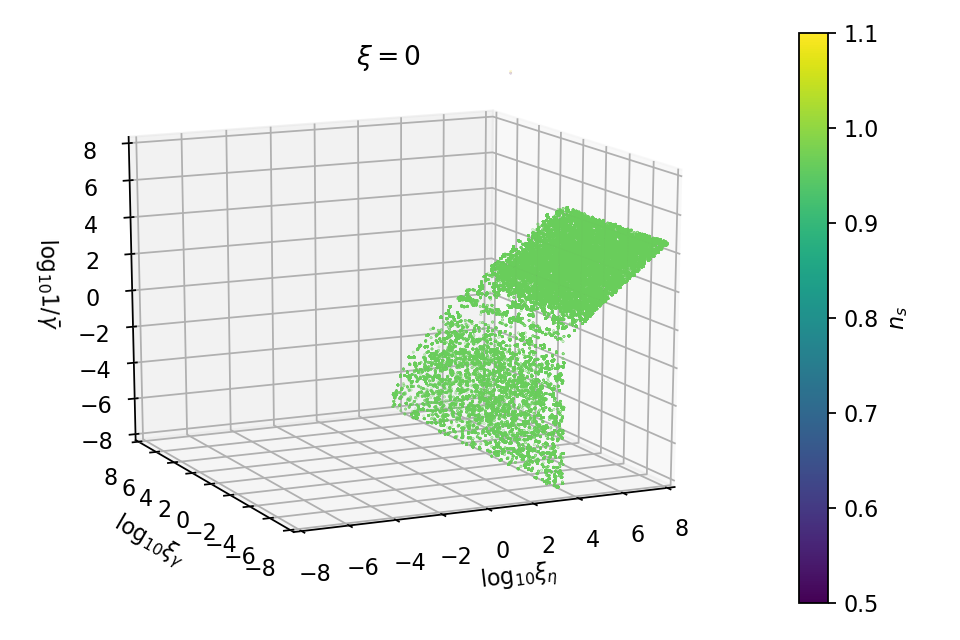} \\(a)}
        \end{minipage}
        \begin{minipage}[h]{0.49\linewidth}
            \center{\includegraphics[width=\textwidth]{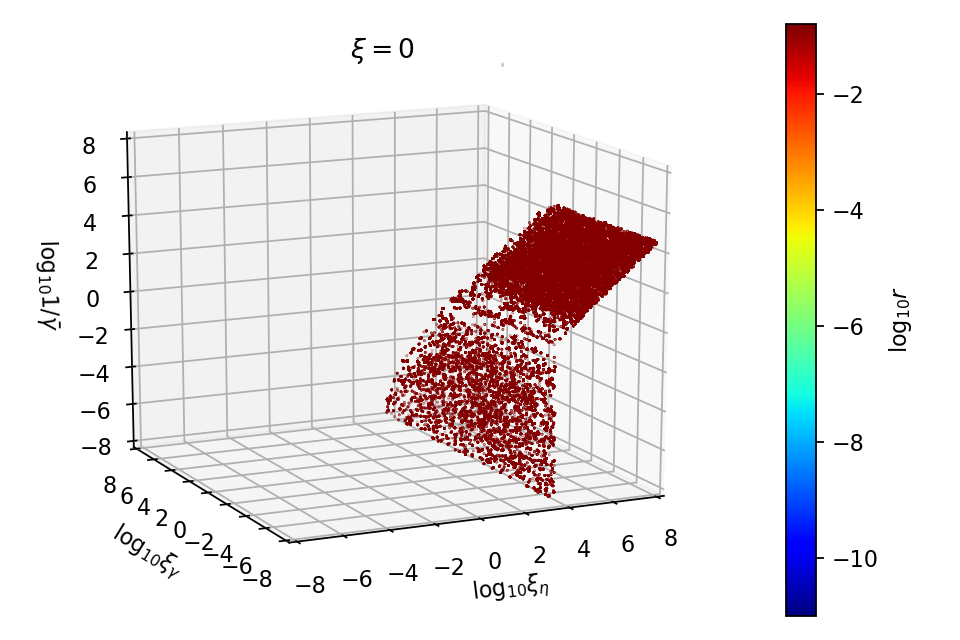} \\(b)}
        \end{minipage}
        \caption{Spectral tilt (a) and tensor-to-scalar ratio (b) in the case $\xi = 0$. }
        \label{fig:Gen4}
    \end{center}
\end{figure}

In this section we study the full parameter space of the theory~\eqref{S_eff} with the four couplings~\eqref{params} taking general positive values. The main result of our scan is that successful inflation is possible in a significant portion of the parameter space.

We implement the numerical procedure described in appendix~\ref{sec:numerical_procedure}. Briefly, it amounts to the following. First, we determine the region of the parameter space where the observed CMB normalization \eqref{CMB_norm} is obtained at the number of e-foldings given in \Eq (\ref{N}). Generally, this region defines a three-dimensional hypersurface. Then we compute $n_s$ and $r$ within this region and reject the points leading to $n_s< 0.5$. There are specific corners of the parameter space where one or multiple periods of brief violation of slow-roll can take place. We have verified that the contribution of this periods to the total $N_\star$ does not exceed $1.1$. It will be interesting to see if such behaviour can lead to observable features.

Figures~\ref{fig:NY} and~\ref{fig:Holst} represent two-dimensional slices of the hypersurface defined by \Eqs \eqref{CMB_norm} and (\ref{N}). In order to show how inflationary observables depend on the parameters of the theory, we present also three-dimensional slices of the hypersurface. This slices are taken by fixing one out of the four couplings \eqref{params} to a specific value. 

We consider four cases: $\xi_\gamma=0$, $\xi_\eta=0$, $\bar{\gamma}=1$, and $\xi = 0$. They generalize the particular corners of the parameter space studied analytically in the previous sections. For example, setting $\bar{\gamma}=1$ allows us to study inflation in the regime when both Einstein-Hilbert and Holst terms are equally important. In each case we scan over the remaining three couplings and select those satisfying the CMB normalization and $n_s>0.5$. Our results are shown in figures~\ref{fig:Gen1}--\ref{fig:Gen4}. First, one observes that the results presented in figures \ref{fig:NY} and \ref{fig:Holst} are reproduced. Second, we see that the theory (\ref{S_eff}) allows for inflation compatible with observations for a broad range of the parameters (\ref{params}).

Specifically, in figure~\ref{fig:Gen1} we consider the case $\xi_\gamma = 0$. The lower part of the plot reproduces Nieh-Yan inflation, see figure~\ref{fig:NY}. Notice that the transition region between the large and small values of $\xi$ is less pronounced due to a lower density of points in the three-dimensional scan. There is a large surface corresponding to metric Higgs inflation at $\xi \simeq 10^3$. Another surface at $\xi \lesssim 10^2$ is ruled out because of the large tensor-to-scalar ratio $r\simeq 0.1$. The region between the two surfaces is interesting since it contains values of $r$ that are potentially detectable in the near future. Figure~\ref{fig:Gen2} shows the case $\xi_\eta = 0$. In the large plane in the front, it reproduces figure~\ref{fig:Holst}. The transition region is again sampled very sparsely. In figure~\ref{fig:Gen3} we consider a more general case $1/ \bar{\gamma} = 1$, where all three non-minimal couplings are important. The region of small $\xi$ yields large values of $r$ which are ruled out by observations. We remark that the region at the bottom resembles figure~\ref{fig:NY}, with $\xi_\eta$ replaced by $\xi_\gamma$. Finally, we consider the case $\xi = 0$ in figure~\ref{fig:Gen4}. Surprisingly, the values of $n_s$ are compatible with observations. However, the minimal value of the tensor-to-scalar ratio found in our study with $\xi = 0$ 
is $r\simeq0.14$. Thus, we conclude that a vanishing non-minimal coupling $\xi$ is not viable.

Figures~\ref{fig:Gen1}--\ref{fig:Gen4} demonstrate that the theory~\eqref{S_eff} describes a very rich and interesting set of inflationary models. What is more, a significant part of these models is in agreement with the observational bounds. In particular, we observe that for most of the parameter choices which we consider, the spectral index is very close to 
\begin{equation}
	n_s = 1 - \frac{2}{N_\star} \,.
\end{equation}
For this reason, a great majority of points in the $n_s$-plots are displayed with the same color.

\section{Validity of classical analysis}
\label{sec:fermions}

\subsection{Estimation of cutoff}
\label{ssec:cutoff}

The goal of this section is to investigate the robustness of inflation in the theory (\ref{S_eff}). We restrict ourselves to analytical analysis and leave a more complete study to future work. The analytical treatment is possible in the cases of Nieh-Yan and Holst inflation studied in sections \ref{sec:NiehYan} and \ref{sec:Holst} respectively.

As a first step, let us estimate the perturbative cutoff $\Lambda$ in the scalar sector of the theory (\ref{S_eff}). In general, $\Lambda$ depends on the value of the background field. Following \cite{Bezrukov:2010jz}, we replace $\chi\to\chi+\delta\chi$ in the potential $U(\chi)$ and treat $\chi$ as the background field and $\delta\chi$ as a perturbation. Expanding $U(\chi)$ in powers of $\delta\chi$, one gets a series of operators of dimension $n>4$; the cutoff due to such an operator is
\begin{equation}\label{Lambda_n}
\Lambda_n(\bar{\chi})=\left(\frac{\diff^n U}{\diff\chi^n}\Big|_{\chi=\bar{\chi}}\right)^{-\frac{1}{n-4}} \;,
\end{equation}
and the overall cutoff is $\Lambda(\bar{\chi}) = \min_n \Lambda_n(\bar{\chi})$.

First, we estimate the cutoff at low energies, \ie for the vanishing background field. To this end, we use that $\diff/\diff \chi = 1/\sqrt{K(h)} \diff/\diff h$ and differentiate repeatedly. In the case of Nieh-Yan inflation, we can use the fact that $\xi\gg1$ as well as $\xi_\eta\gg1$. Then, the higher-dimensional operators are suppressed by the scale
\begin{equation}\label{CutoffNiehYan}
	\Lambda_{\rm NY}(0)=\frac{M_P}{\sqrt{c_1 \xi + c_2 \xi_\eta^2}} \,,
\end{equation}
where $c_1$ and $c_2$ are coefficients of the order of one. For $\xi_\eta\lesssim \sqrt{\xi}$, this leads to $ \Lambda_{\rm NY}(0)\sim M_P/\sqrt{\xi}$, which agrees with the low-energy cutoff of Palatini Higgs inflation \cite{1012.2900}. In the case $\xi_\eta\gtrsim\sqrt{\xi}$, we get $\Lambda_{\rm NY}(0)\sim M_P/\xi_\eta$. After identifying $\xi_\eta$ with $\xi$, this reproduces the well-known result \cite{0902.4465,0903.0355} for the metric case.

One can proceed similarly to find the perturbative cutoff in a vacuum background in the case of Holst inflation. We obtain that the scale of higher-order operators is
\begin{equation}\label{CutoffHolst}
\Lambda_{\rm Holst}(0) \sim \begin{cases}
\frac{M_P}{\sqrt{\xi}} & \frac{1}{\bar{\gamma}} \lesssim \frac{1}{\sqrt{\xi}} \\
\frac{M_P\bar{\gamma}}{\xi}  &\frac{1}{\sqrt{\xi}}\lesssim \frac{1}{\bar{\gamma}} \lesssim 1 \\
\frac{M_P}{\xi} & 1 \lesssim \frac{1}{\bar{\gamma}} 
\end{cases} \,.
\end{equation}
In agreement with our previous analysis, we conclude that only for $1/\bar{\gamma}\lesssim 1/\sqrt{\xi}$, we fully reproduce Palatini Higgs inflation. For higher values of $1/\bar{\gamma}$, the cutoff is lowered. This matches the fact that part of the inflationary or postinflationary potential corresponds to the metric Higgs case. Let us stress again that, although in the parameter range $1/\sqrt{\xi} \lesssim 1/\bar{\gamma} \lesssim N_\star$ the CMB perturbations are generated during the Palatini phase, the dynamics at smaller field values and, in particular, during preheating, is different from that in Palatini Higgs inflation. Moreover, the lower cutoff likely implies that the connection of low- and high-energy physics, which may be present in Palatini Higgs inflation \cite{Shaposhnikov:2020fdv}, is lost. We solely consider higher-order operators generated in the Higgs sector of the theory, thus the full analysis of the cutoff in Holst inflation remains to be done.
  
Now we turn to the cutoff in an inflationary background. In the case of Nieh-Yan inflation, we use the potential in the form \eqref{NY_PotInfl} and obtain that
 \begin{equation}
 \frac{\diff^n U}{\diff\chi^n}\Big|_{\chi=\bar{\chi}} \sim \frac{\lambda M_P^4}{\xi^2}\exp\left(-\frac{2\xi}{\sqrt{\xi+6\xi_\eta^2}}\frac{\bar{\chi}}{M_P}\right)\left(\frac{\xi^2}{M_P^2(\xi+6\xi_\eta^2)} \right)^{\frac{n}{2}} \;.
 \end{equation}
 It follows that the lowest scale is achieved for $n\rightarrow \infty$ and the cutoff during inflation is
\begin{equation}
    \Lambda_{\rm inf,\: NY}=M_P\sqrt{\frac{\xi+6\xi_\eta^2}{\xi^2}} \;.
\end{equation}
Since we have $\Lambda_{\rm inf,\: NY}> M_P/\sqrt{\xi}$, it always lies above the scale $\mu_{\rm inf}$ of inflation, as it should. The latter is given by $y_t M_P/\sqrt{\xi}$, where $y_t$ is the inflationary value of the top Yukawa coupling \cite{Shaposhnikov:2020fdv}. The Hubble scale $H\sim M_P/\xi$ is even lower. We must emphasize, however, that a complete analysis of the perturbative cutoff scale in Nieh-Yan inflation remains to be done, once the rest of the Standard Model degrees of freedom are included into consideration. In particular, taking into account longitudinal modes of gauge bosons is expected to lower the inflationary cutoff, as it does in the metric case \cite{Bezrukov:2010jz}.

\subsection{Fermion energy density during inflation: classical analysis}
\label{ssec:class}

Thanks to the presence of torsion, the effective metric theory (\ref{S_eff}) includes five- and six-dimensional operators containing interaction between the Higgs field and the vector and axial fermion currents. We consider first the scalar-fermion interaction given in the second and the third lines of \Eq (\ref{S_hVA}). The dominant contribution to the interaction energy density comes from the $g^{00}$-term which contains the time derivative of the background field $h$. The $0^\text{th}$ components of the fermion currents can be estimated as $V_0\sim\mu_V^3$, $A_0\sim\mu_A^3$, where $\mu_{V,A}$ are chemical potentials corresponding to the vector and axial currents, correspondingly.\footnote
{The contributions due to the Hubble temperature, $V_0\sim A_0 \sim H^3$, turn out to be smaller.}
In turn, the chemical potentials at zero temperature are equal to the coefficients of the currents. We can write
\begin{equation}\label{muVA}
\mu_{V,A}=\frac{3\dot{\Omega}}{2\Omega}b_{V,A}+\frac{3}{4(\gamma^2+1)}\left(\frac{\dot{\bar{\eta}}}{\Omega^2}+\dot{\gamma} \right)c_{V,A} \;,
\end{equation}
where dot denotes the time derivative and
\begin{equation}\label{bbcc}
b_V=\alpha \;, ~~~ b_A=\beta \;, ~~~ c_V=\alpha\gamma \;, ~~~ c_A=1+\beta\gamma \;.
\end{equation}
The scalar-fermion interaction energy density is estimated as
\begin{equation}
V_{V,A}\sim \mu_{V,A}^4 \;.
\end{equation}
This needs to be compared with the inflationary energy density $U$ determined by \Eq (\ref{PotChi}). 

Let us estimate the relevance of the higher-order fermionic interactions for Nieh-Yan inflation. One finds readily that $c_V=0$ and $c_A=1$. Moreover, we have $\dot{\gamma}=0$ and $\dot{\Omega}$, $\dot{\bar{\eta}}$ are proportional to $\dot{h}=-\sqrt{2\epsilon U(h)/(3K(h))}$. Hence, the largest value of $\dot{h}$ is achieved at the end of inflation.
Using \Eqs (\ref{NY_eps}), (\ref{NY_hend}) and (\ref{GammaEta}) as well as the fact that during inflation $\xi h^2\gg M_P^2$, we obtain
\begin{equation} \label{fermionChemicalPotential_NY}
\mu_V^2\lesssim \frac{3\lambda\alpha^2 M_P^2}{64\xi^2} \;, ~~~ \mu_A^2\lesssim\frac{3\lambda (\xi_\eta+\beta\xi)^2  M_P^2}{64\xi^4} \;.
\end{equation}
Consider first the case of vanishing non-minimal couplings, $\alpha=\beta=0$. Demanding that $\mu_{V,A}^4 \ll U=\lambda M_P^4/(4\xi^2)$, from eq.~(\ref{fermionChemicalPotential_NY}) we obtain
\begin{equation} \label{fermion_NYCondition}
\frac{\xi^6}{\xi_\eta^4}\gg 10^{-2}\lambda \;.
\end{equation}
In section \ref{sec:NiehYan} we showed that $\xi_\eta \lesssim \xi$ and $\xi\gg 1$ in all phenomenologically acceptable scenarios. Therefore, the condition \eqref{fermion_NYCondition} is always fulfilled for the values of the Higgs self-coupling typical for inflation, \Eq (\ref{Lambda}). In particular, this is true in the metric and Palatini limits. 

From eq.~(\ref{fermionChemicalPotential_NY}) we see that the condition \eqref{fermion_NYCondition} is also valid for non-zero $\beta$ as long as $\beta^2\lesssim\xi^2_\eta/\xi^2$. For general values of non-minimal couplings we can view the requirement $\mu_{V,A}^4\ll U$ as a condition on $\alpha$, $\beta$ and obtain
\begin{equation}\label{ab_bound1}
\alpha^2,~\beta^2 \ll \frac{10\xi}{\lambda^{1/2}} \;.
\end{equation}
Using again the typical inflationary value of $\lambda$, \Eq (\ref{Lambda}), we see that $\alpha^2$, $\beta^2$ are bounded from above by $\sim \xi\gg 1$.

Now we estimate the contribution from the fermion current interaction terms written in the second line of \Eq (\ref{S_hVA}). We have
\begin{equation}\label{V_VVVAAA}
	V_{VV,VA,AA}=\frac{3}{16M_P^2(\gamma^2+1)}\left((\beta^2-1-2\beta\gamma)\mu_A^6+2\alpha(\beta-\gamma)\mu_A^3\mu_V^3+\alpha^2\mu_V^6\right) \;.
\end{equation}
In Nieh-Yan inflation we use \Eq \eqref{fermionChemicalPotential_NY} for the chemical potentials. Consider first the vanishing non-minimal couplings. Then, requiring $V_{VV,VA,AA}$ to be much smaller than the inflationary density gives
\begin{equation}
\frac{10^4\xi^{10}}{\lambda^2\xi_\eta^6}\gg 1 \;.
\end{equation}
This bound is weaker than the one due to the scalar-fermion interaction, \Eq \eqref{fermion_NYCondition}. It is also valid for non-zero $\beta$ as long as $\beta^2\lesssim 1$. For general values of non-minimal couplings, we obtain again the condition (\ref{ab_bound1}).

Let us now discuss the Holst inflation. We find that $c_V=\alpha/(\bar{\gamma}\Omega^2)$ and $c_A=1+\beta/(\bar{\gamma}\Omega^2)$. We assume that the strongest constraint on the parameters comes from the end of inflation.\footnote{We do not consider the possible transitory region between the metric and Palatini regimes, since in this region the slow-roll condition is violated anyways.} The form of the constraint depends on the value of the Barbero-Immirzi parameter. Indeed, as discussed in section \ref{sec:Holst}, depending on whether $\bar{\gamma}\lesssim 1 $ or $\bar{\gamma}\gtrsim 1$, the end of inflation happens as in the metric $h_{\rm end}\sim M_P/\sqrt{\xi}$ or Palatini $h_{\rm end}\sim M_P$ scenario, correspondingly. Repeating the analysis made above for the Nieh-Yan inflation, we arrive at the following expressions for the chemical potentials:
\begin{equation}
	\mu_V^2\lesssim\left\lbrace\begin{array}{ll}\displaystyle \frac{49 \lambda \alpha^2  \bar{\gamma}^4 M_P^2}{108 \xi^2} \;, & \displaystyle\bar{\gamma}\lesssim 1\\\\
	\displaystyle	\frac{3\lambda\alpha^2 M_P^2}{64\xi^2}  \;, & \displaystyle \bar{\gamma}\gtrsim 1 \end{array}\right. \;, ~~~\displaystyle \mu_A^2\lesssim\left\lbrace\begin{array}{ll}\displaystyle \frac{49 \lambda \bar{\gamma}^4(\beta-\frac{3}{7\bar{\gamma}})^2 M_ P^2}{108 \xi^2} \;, &\displaystyle \bar{\gamma}\lesssim 1\\\\
	\displaystyle	\frac{3\lambda(\beta-\frac{1}{8\bar{\gamma}\xi})^2 M_P^2}{64\xi^2} \;, &\displaystyle \bar{\gamma}\gtrsim 1 \end{array}\right. \;,
\end{equation}
where we made use of the fact that $\xi\gg 1$ (see figure \ref{fig:Holst}). Note that numerical prefactors can change by order $1$ depending on how the end of inflation is defined.
Now we consider the constraint due to the scalar-fermion interaction terms. In the case of zero fermion non-minimal couplings, $\alpha=\beta=0$, the requirement $\mu_{V,A}^4 \ll U$ reduces to the conditions
\begin{align}
	& \frac{\xi^2}{\bar{\gamma}^4}\gg 10^{-2}\lambda \;, ~~~ \bar{\gamma}\lesssim 1\;, \\
	& \bar{\gamma}^4\xi^6\gg 10^{-6}\lambda \;, ~~~ \bar{\gamma}\gtrsim 1\;,
\end{align}
which are satisfied for the phenomenologically viable values of the Higgs non-minimal coupling. For generic $\alpha$ and $\beta$, we get the conditions
\begin{equation}
	\alpha^2,\beta^2\ll \left\lbrace\begin{array}{ll}\displaystyle \frac{\xi}{\lambda^{1/2}\bar{\gamma^4}} \;, & \displaystyle \bar{\gamma}\lesssim  1\;, \\\\
\displaystyle	\frac{10\xi}{\lambda^{1/2}}\;, &\displaystyle \bar{\gamma}\gtrsim  1 \;.\end{array}\right.
\end{equation}
We see that in the Palatini region, \ie for sufficiently large $\bar{\gamma}$, the bound on $\alpha$ and $\beta$ agrees with the corresponding result in the Nieh-Yan case, \Eq \eqref{ab_bound1}.
Finally, considering the fermion-fermion interaction terms (\ref{V_VVVAAA}) shows that the corresponding constraints are even weaker than the ones above.

\subsection{Fermion energy density during inflation: quantum corrections}
\label{ssec:quant}

The fermion current interaction terms also lead to a quantum contribution to the effective potential. For example, we can consider the two-loop diagram
\begin{equation}\label{Loop}
\begin{minipage}[h]{0.1\linewidth}
\center{\includegraphics[width=0.5\linewidth]{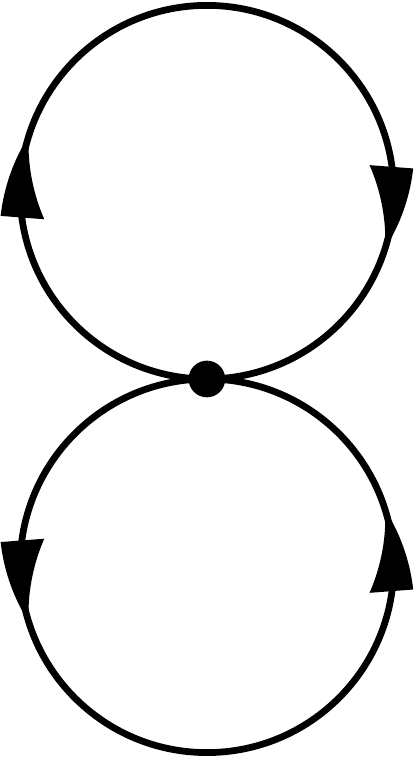}}
\end{minipage}\sim\frac{\varkappa}{M_P^2}\frac{(3\cdot 4)^2}{(16\pi^2)^2}\mu_{\rm inf}^6 \;,
\end{equation}
where we took into account the degeneracy due to color and spin.
Moreover, $\varkappa$ is the coefficient of the respective current-current interaction and $\mu_{\rm inf}$ corresponds to the relevant inflationary energy scale. The latter can be evaluated via the top quark mass during inflation (see, e.g., \cite{Shaposhnikov:2020fdv}):
\begin{equation}\label{Mu}
\mu_{\rm inf}=y_t\frac{M_P}{\sqrt{\xi}} \;,
\end{equation}
where $y_t$ is the top quark Yukawa coupling and we made use of \Eq (\ref{PotChi}). The quantity (\ref{Loop}) needs to be much smaller than the energy density $U=\lambda M_P^4/(4 \xi^2)$ of the classical inflationary background. Using \Eqs (\ref{Loop}) and (\ref{Mu}), we obtain the following bound on $\varkappa$:
\begin{equation}\label{Kappa}
|\varkappa|\ll\frac{4\pi^4\lambda\xi}{9 y_t^6} \;.
\end{equation}
Note that both $\lambda$ and $y_t$ are running couplings, and in the above condition they must be evaluated at the inflationary energy scale. For the Yukawa coupling we adopt the following value \cite{Shaposhnikov:2020fdv}:
\begin{equation}
y_t=0.43 \;.
\end{equation}
We keep the inflationary value of $\lambda$ as free parameter.

The condition (\ref{Kappa}) can be converted into conditions on the couplings $\lambda$, $\alpha$, $\beta$, $\bar{\gamma}$, $\xi$ and $\xi_\gamma$. We make an analytical estimate in the particular cases of Nieh-Yan and Holst inflation. In the former scenario, the vector-vector, axial-vector and axial-axial interaction terms appear in the effective action (\ref{S_hVA}) with the following coefficients: 
\begin{align}
& \varkappa_{VV}=\frac{3}{16}\alpha^2 \;, \\
& \varkappa_{AV}=\frac{3}{8}\alpha\beta \;, \\
& \varkappa_{AA}=\frac{3}{16}(\beta^2-1) \;.
\end{align}
In the absence of non-minimal fermion couplings, $\alpha=\beta=0$, \Eq (\ref{Kappa}) becomes the condition on the Higgs quartic coupling,
\begin{equation}\label{quantumPotentialCondition_NY2}
\lambda \xi \gg 10^{-4} \;,
\end{equation}
which is fulfilled in all phenomenologically acceptable scenarios. In the general case and if no fine-tuning among the couplings is assumed, we obtain
\begin{equation} \label{quantumPotentialCondition_NY}
10^{4} \lambda\xi \gg \max(1,\alpha^2,\beta^2)\;.
\end{equation}
For typical inflationary values $\lambda \sim 10^{-3}$, we roughly obtain $\max(\alpha^2,\beta^2)<\xi$. It is important to note that the conditions (\ref{quantumPotentialCondition_NY2}), (\ref{quantumPotentialCondition_NY}) do not depend on $\xi_\eta$. Thus, they are also applied to the case of pure Palatini Higgs inflation, $\xi_\eta =0$.

In the case of Holst inflation, the coefficients in the current-current interaction terms are field-dependent; they are given by
\begin{align}
& \varkappa_{VV}=\frac{3}{16}\frac{\alpha^2\bar{\gamma}^2\Omega^4}{\bar{\gamma}^2\Omega^4+1} \;, \\
& \varkappa_{AV}=\frac{3}{8}\frac{\alpha\bar{\gamma}^2\Omega^4}{\bar{\gamma}^2\Omega^4+1}\left(\frac{1}{\bar{\gamma}\Omega^2}-\beta\right) \;, \\
& \varkappa_{AA}=\frac{3}{16}\frac{\bar{\gamma}^2\Omega^4}{\bar{\gamma}^2\Omega^4+1}\left(\beta^2-1-\frac{2\beta}{\bar{\gamma}\Omega^2}\right) \;.
\end{align}
We assume again that the strongest constraint on the coefficients comes from the end of inflation. In the absence of non-minimal fermion couplings, $\alpha=\beta=0$, we obtain the combined condition 
\begin{equation}\label{quantumPotentialCondition_Holst}
10^4\lambda\xi\gg \frac{\bar{\gamma}^2}{\bar{\gamma}^2+1} \;,
\end{equation}
which is satisfied for all parameters suitable for phenomenology. For generic $\alpha$ and $\beta$ we get
\begin{equation}\label{quantumPotentialCondition_Holst2}
10^{4} \lambda\xi \gg \max(1,\frac{|\alpha|}{\bar{\gamma}},\frac{|\beta|}{\bar{\gamma}},\alpha^2,\beta^2)\cdot\frac{\bar{\gamma}^2}{\bar{\gamma}^2+1} \;.
\end{equation}
Note that in the Palatini limit of Holst inflation $\bar{\gamma}\gg 1$, and the constraints (\ref{quantumPotentialCondition_Holst}), (\ref{quantumPotentialCondition_Holst2}) coincide with \Eqs (\ref{quantumPotentialCondition_NY2}), (\ref{quantumPotentialCondition_NY}).

\section{Discussion and outlook}
\label{sec:conclusion}

In this paper, we discussed Higgs inflation in the Einstein-Cartan (EC) formulation of gravity. Working within the EC framework, one can limit oneself to the same gravitational action as in the metric \cite{Bezrukov:2007ep} and Palatini \cite{Bauer:2008zj} Higgs inflation: the Einstein-Hilbert term with a non-minimal coupling to the Higgs boson. If one furthermore disregards fermions, then it is well-known that EC theory reduces to the Palatini version of gravity. However, the EC case allows for a more general gravitational action including the Holst term and the Nieh-Yan term, both coupled to the Higgs field. This action is displayed in \Eq \eqref{S_grav} and contains three more independent couplings than in the Palatini case. 

We analyzed inflation in the resulting theory analytically in the two special cases. The first one corresponds to neglecting the Holst term and the non-minimal coupling to it. Then only the coupling of the Higgs field to the Nieh-Yan term is present in addition to the standard gravitational action of Higgs inflation. We dubbed this scenario ``Nieh-Yan" inflation. We showed that it is consistent with observations in a broad range of parameters. Moreover, it encompasses metric and Palatini Higgs inflation, which are reproduced in suitable regions of the parameter space, and allows for a continuous interpolation between them. The second scenario considered analytically amounts to neglecting the non-minimal couplings both to the Holst and Nieh-Yan terms. Then only the Holst action is left in addition to the standard gravitational action, and we called this ``Holst inflation''. Again, we found that large parts of the parameter space reproduce the predictions of either metric or Palatini Higgs inflation. However, unlike the Nieh-Yan case, the region interpolating between the two models is not compatible with observations. 

As a next step, we studied the whole parameter space of the theory numerically. In general, the requirement that the observed amplitude of CMB perturbations is reproduced fixes one parameter and leaves unconstrained the other three. We restricted ourselves to positive values of the couplings and found that inflation can successfully take place in virtually all parts of the parameter space. Moreover, it turns out that the spectral index $n_s$ is mostly independent of the choice of couplings and lies very close to $1-2/N_\star$. In contrast, the tensor-to-scalar ratio $r$ can vary, roughly, between $1$ and $10^{-10}$. This allows us to rule out some corners of the parameter space where $r$ exceeds the observational bound. What is more interesting, we found regions where $r$ is slightly below the current bound, in which case primordial gravitational waves become potentially detectable in the near future. 

Up to this point, our analysis was classical. In the end, we gave a short discussion of quantum effects for the special cases of Nieh-Yan and Holst inflation. First, we studied the cutoff scale, above which perturbation theory breaks down. Secondly, we investigated the influence of additional higher-dimensional fermion-fermion and Higgs-fermion interactions that arise in EC gravity. In both cases, we concluded that quantum effects only lead to small corrections and do not invalidate our analysis. However, their complete study in the full parameter space of the theory remains to be done. As an outlook, it would be very interesting to take into account the running of the Higgs quartic coupling, since the latter affects greatly the inflationary potential, leading to a bunch of different scenarios for inflation. Whether this running can be computed within the Standard Model is an open question and likely depends on the choice of parameters in the EC action. Finally, we remark that we relied on the slow-roll approximation throughout our study. Since we observed that it can be briefly violated, an analysis beyond this approximation remains to be done.

In this paper, we studied the generalization of Palatini Higgs inflation to the EC case. Previously \cite{Shaposhnikov:2020geh}, we have proposed that the Palatini formulation of gravity may serve another purpose, namely to address the question of why the Electroweak scale $v$ is so much smaller than the Planck scale $M_P$. Based on earlier works \cite{Shaposhnikov:2018xkv, Shaposhnikov:2018jag,Shkerin:2019mmu}, we developed a scenario in which the Standard Model is classically scale-invariant, \ie the Higgs mass is put to zero. In addition, we assumed that no new degrees of freedom appear above $v$. In this framework, we suggested that a non-perturbative gravitational effect can be responsible for generating $v$ out of $M_P$ via a gravitational-scalar instanton. Schematically,
\begin{equation}\label{hierarchy}
v\sim M_Pe^{-\mathcal{W}} \;,
\end{equation} 
where $\mathcal{W}$ is the instanton action. Invariably, the instanton action is mostly sensitive to the coefficient of the kinetic term of the Higgs field in the limit of large field values or, specifically, to the quadratic residue of $K(h)$ at zero in that limit \cite{Karananas:2020qkp}.
If we denote by $\kappa$ the asymptotics of $M_P^2/(h^2K(h))$ at large $h$, then \cite{Shaposhnikov:2018xkv,Karananas:2020qkp}
\begin{equation}
\mathcal{W}=\sqrt{\kappa}\:\tilde{\mathcal{W}} \;,
\end{equation}
where $\tilde{\mathcal{W}}$ depends on $\kappa$ at most logarithmically. From here one can expect that big values of $\mathcal{W}$, which are necessary for generating the small ratio $v/M_P$, are generally achieved when $\kappa\gg 1$. Let us apply this argument to the Nieh-Yan inflation. From \Eq (\ref{NY_K}) we have
\begin{equation}
\kappa=\frac{\xi^2}{\xi+6\xi_\eta^2} \;.
\end{equation}
Thus, as soon as $\xi_\eta\ll \sqrt{\xi}$, one can expect the non-perturbative mechanism to work successfully. In particular, at $\xi_\eta=0$ the Palatini Higgs inflation model is reproduced, in which it was checked that the gravitational-scalar instanton can yield the hierarchy (\ref{hierarchy}) \cite{Shaposhnikov:2020geh}. It would be very interesting to study this mechanism of generating the weak scale from gravity in the general case of EC theory.

One can go one step further and also remove the dimensionful parameter of gravity, the Planck mass $M_P$. This can be achieved by introducing an extra degree of freedom -- the scalar dilaton. In a no-scale scenario, dimensionful parameters, such as the Planck mass and the Higgs vacuum expectation value, are replaced by the dilaton field with appropriate coupling constants. The Planck mass is generated dynamically as a result of spontaneous breaking of the scale symmetry by the dilaton. In metric gravity, a concrete example of a phenomenologically viable theory comprising the Standard Model and General Relativity in the scale-invariant setting is the Higgs-Dilaton model introduced in \cite{Shaposhnikov:2008xb} and studied extensively in \cite{GarciaBellido:2011de,Bezrukov:2012hx}. Recently, an investigation of the Higgs-dilaton model in the Palatini formulation of gravity was initiated in \cite{Rubio:2020zht, Shaposhnikov:2020frq}. An interesting task for future work would be to generalize these models to EC theory and study their implications for cosmology.

\appendix
\section{Numerical procedure}
\label{sec:numerical_procedure}

Let us outline the numerical procedure used to obtain figures \ref{fig:NY}, \ref{fig:Holst} and \ref{fig:Gen1}--\ref{fig:Gen4}. It consists of three steps: \emph{(i)} calculating the number of e-foldings $N_\star$ for any values of $\xi,\; \xi_\gamma,\; \xi_\eta,$ and $ \bar{\gamma}$; \emph{(ii)} sampling of the parameter space leading to a certain value of $N_\star$; \emph{(iii)} computation of the cosmological observables $n_s$ and $r$ in the determined regions of the parameter space. 

\textit{Calculating $N_\star$.} The number of the e-foldings reads as follows,\footnote{For the reader's convenience we repeat some formulas from the main text here.}
\begin{equation}
    N_\star=\frac{1}{M_P^2}\limitint_{h_{\rm end}}^{h_\star} dh K(h) \frac{U(h)}{\frac{\diff U(h)}{\diff h}}\,,
    \label{N_ef}
\end{equation}
with the potential given by 
\begin{equation}
U(h)=\frac{\lambda h^4}{4\Omega(h)^4}
\end{equation}
and the kinetic term function defined in \Eq \eqref{modifiedK}. The upper integration limit is fixed  due to the CMB-normalization (cf \Eq \eqref{CMB_norm})
\begin{equation}
    U(h_\star)/\epsilon(h_\star) = 5.0 \cdot 10^{-7} M_P^4.
    \label{CMB_norm2}
\end{equation}
 Next, the slow roll parameters \Eqs \eqref{NY_eps} and \eqref{NY_eta} can in general be computed analytically, and the lower limit in~\eqref{N_gen} is given by the condition that either $\epsilon$ or $|\eta|$ reaches $1$. Thus, the procedure for calculating $N_\star$ for given values of $\xi,\; \xi_\gamma,\; \xi_\eta,$ and $ \bar{\gamma}$ is the following:
\begin{itemize}
	\item One finds $h_\star$ solving \Eq \eqref{CMB_norm2}.
	\item One determines the end of inflation by solving $\epsilon(h_{\rm end})\simeq 1$
	and $|\eta(h_{\rm end})|\simeq 1$ and requiring that at least one of the slow-roll parameters remain larger than $1$ for all smaller values of the field.
	The largest of two values of $h_{\rm end}$ should be selected. A certain care must be taken at this step to make sure that the possible short periods of the slow-roll violation (cf. figure~\ref{fig:HolstTransition}) are not misinterpreted as the and of inflation. Indeed, even though $|\eta|$ can be larger than $1$ during these periods, their contribution to \Eq \eqref{N_ef} is at the percent level.
	\item Once the limits have been determined, the integral in \Eq \eqref{N_ef} is evaluated numerically.
\end{itemize}

\textit{Sampling of the parameter space.} We can use ${N_\star=N_\star(\xi,\; \xi_\gamma,\; \xi_\eta, \; \bar{\gamma})}$ to study the parameter space of the theory (\ref{S_eff}). We fix the value of $N_\star$ as in \Eq (\ref{N}). This condition defines a three-dimensional surface in the four-dimensional parameter space. In sections~\ref{sec:NiehYan}, \ref{sec:Holst} and \ref{sec:gen} we present the two- and three-dimensional slices of this surface.

In order to efficiently sample the parameter space we use \texttt{emcee} sampler which employs the ensemble Markov chain Monte Carlo (MCMC) method, see~\cite{ForemanMackey:2012ig} and references therein. The log-likelihood function for the sampler is given by
\begin{equation}
\log L = -\frac{(55 - N_\star(\xi,\; \xi_\gamma,\; \xi_\eta, \; \bar{\gamma}))^2}{\sigma_N^2} \;,
\end{equation}
with $\sigma_N = 0.01$. We choose flat priors for logarithms of the non-minimal couplings. Note that we employ MCMC technique -- which is commonly used for Bayesian inference -- as a mere numerical tool allowing for efficient sampling. Thus we do not assign any special meaning to $\sigma_N$. 
At the end we select the points in the parameter space which result in $N_\star$ being sufficiently close to $55$.

\textit{Computation of the observables.} Once the parameter space is sampled, we determine the cosmological parameters through $n_s=1-6\epsilon+2\eta$, $r=16\epsilon$. Not all points satisfying condition~\eqref{N} lead to reasonable values of $r$ and $n_s$. In fact, for some of them $n_s$ can be very small or even negative. We discard all parameter sets for which $n_s<0.5$.

\acknowledgments

The work was supported by ERC-AdG-2015~grant~694896 and by the Swiss National Science Foundation Excellence grant~200020B\underline{ }182864.

\paragraph{Note added.} While the present work was in preparation, a paper which studies similar and complementary questions appeared \cite{Langvik:2020nrs}. The scalar part of the action, which results in the non-minimal kinetic term \eqref{modifiedK}, coincides with our result (with the identifications $H_0=1/\bar{\gamma}$, $H_1 = \xi_\gamma/\bar{\gamma}$ and $Y_1 = - \xi_\eta$). Whereas the analytic study of inflation also largely agrees, \cite{Langvik:2020nrs} does not discuss the possibility of an intermediate slow-roll violation or the critical value of $1/\bar{\gamma}$ (see \Eq \eqref{borderGamma}). 
	 The choice of parameters is more general in \cite{Langvik:2020nrs} as the signs of $\xi$, $\xi_\gamma$ and $\xi_\eta$ are not restricted. This allows for inflationary solutions with $\xi=0$ and $\xi_\gamma<0$ \cite{Langvik:2020nrs}. If like in our study one only considers $\xi_\gamma>0$, no observationally acceptable solutions with $\xi=0$ exist.
	 The numerical investigation of \cite{Langvik:2020nrs} seems to rely on a similar method as ours. However, the presentation of results is different. In particular, it is clear from our plots that the spectral index is given by $n_s=1-2/N_\star$ in most parts of parameter space. No fermions or quantum effects are considered in \cite{Langvik:2020nrs}.

\paragraph{Statement of provenance.} This document is based on the Accepted Manuscript version of an article accepted for publication in Journal of Cosmology and Astroparticle Physics. Neither SISSA Medialab Srl nor IOP Publishing Ltd is responsible for any errors or omissions in this version of the manuscript or any version derived from it.  The Version of Record is available online at \href{https://iopscience.iop.org/article/10.1088/1475-7516/2021/02/008}{\emph{JCAP} {\bfseries 02} (2021) 008} (and erratum at \href{https://iopscience.iop.org/article/10.1088/1475-7516/2021/10/E01}{\emph{JCAP} {\bfseries 10} (2021) E01}).

\bibliographystyle{JHEP}
\bibliography{EC}

\providecommand{\href}[2]{#2}\begingroup\raggedright\begin{thebibliography}{10}

\bibitem{BENNETT1993409}
C.~Bennett, N.~Boggess, E.~Cheng, M.~Hauser, T.~Kelsall, J.~Mather et~al.,
  \emph{Scientific results from cobe},
  \href{https://doi.org/https://doi.org/10.1016/0273-1177(93)90150-A}{\emph{Advances
  in Space Research} {\bfseries 13} (1993) 409 }.

\bibitem{1001.4538}
{\scshape WMAP} collaboration, \emph{{Seven-Year Wilkinson Microwave Anisotropy
  Probe (WMAP) Observations: Cosmological Interpretation}},
  \href{https://doi.org/10.1088/0067-0049/192/2/18}{\emph{Astrophys. J. Suppl.}
  {\bfseries 192} (2011) 18} [\href{https://arxiv.org/abs/1001.4538}{{\ttfamily
  1001.4538}}].

\bibitem{Akrami:2018odb}
{\scshape Planck} collaboration, \emph{{Planck 2018 results. X. Constraints on
  inflation}}, \href{https://doi.org/10.1051/0004-6361/201833887}{\emph{Astron.
  Astrophys.} {\bfseries 641} (2020) A10}
  [\href{https://arxiv.org/abs/1807.06211}{{\ttfamily 1807.06211}}].

\bibitem{Starobinsky:1980te}
A.A.~Starobinsky, \emph{{A New Type of Isotropic Cosmological Models Without
  Singularity}}, \href{https://doi.org/10.1016/0370-2693(80)90670-X}{\emph{Adv.
  Ser. Astrophys. Cosmol.} {\bfseries 3} (1987) 130}.

\bibitem{Guth:1980zm}
A.H.~Guth, \emph{{The Inflationary Universe: A Possible Solution to the Horizon
  and Flatness Problems}},
  \href{https://doi.org/10.1103/PhysRevD.23.347}{\emph{Adv. Ser. Astrophys.
  Cosmol.} {\bfseries 3} (1987) 139}.

\bibitem{Linde:1981mu}
A.D.~Linde, \emph{{A New Inflationary Universe Scenario: A Possible Solution of
  the Horizon, Flatness, Homogeneity, Isotropy and Primordial Monopole
  Problems}}, \href{https://doi.org/10.1016/0370-2693(82)91219-9}{\emph{Adv.
  Ser. Astrophys. Cosmol.} {\bfseries 3} (1987) 149}.

\bibitem{Mukhanov:1981xt}
V.F.~Mukhanov and G.V.~Chibisov, \emph{{Quantum Fluctuations and a Nonsingular
  Universe}}, {\emph{JETP Lett.} {\bfseries 33} (1981) 532}.

\bibitem{Bezrukov:2007ep}
F.L.~Bezrukov and M.~Shaposhnikov, \emph{{The Standard Model Higgs boson as the
  inflaton}}, \href{https://doi.org/10.1016/j.physletb.2007.11.072}{\emph{Phys.
  Lett. B} {\bfseries 659} (2008) 703}
  [\href{https://arxiv.org/abs/0710.3755}{{\ttfamily 0710.3755}}].

\bibitem{Bauer:2008zj}
F.~Bauer and D.A.~Demir, \emph{{Inflation with Non-Minimal Coupling: Metric
  versus Palatini Formulations}},
  \href{https://doi.org/10.1016/j.physletb.2008.06.014}{\emph{Phys. Lett. B}
  {\bfseries 665} (2008) 222}
  [\href{https://arxiv.org/abs/0803.2664}{{\ttfamily 0803.2664}}].

\bibitem{Raatikainen:2019qey}
S.~Raatikainen and S.~Rasanen, \emph{{Higgs inflation and teleparallel
  gravity}}, \href{https://doi.org/10.1088/1475-7516/2019/12/021}{\emph{JCAP}
  {\bfseries 12} (2019) 021}
  [\href{https://arxiv.org/abs/1910.03488}{{\ttfamily 1910.03488}}].

\bibitem{Azri:2017uor}
H.~Azri and D.~Demir, \emph{{Affine Inflation}},
  \href{https://doi.org/10.1103/PhysRevD.95.124007}{\emph{Phys. Rev. D}
  {\bfseries 95} (2017) 124007}
  [\href{https://arxiv.org/abs/1705.05822}{{\ttfamily 1705.05822}}].

\bibitem{Rubio:2018ogq}
J.~Rubio, \emph{{Higgs inflation}},
  \href{https://doi.org/10.3389/fspas.2018.00050}{\emph{Front. Astron. Space
  Sci.} {\bfseries 5} (2019) 50}
  [\href{https://arxiv.org/abs/1807.02376}{{\ttfamily 1807.02376}}].

\bibitem{Tenkanen:2020dge}
T.~Tenkanen, \emph{{Tracing the high energy theory of gravity: an introduction
  to Palatini inflation}},
  \href{https://doi.org/10.1007/s10714-020-02682-2}{\emph{Gen. Rel. Grav.}
  {\bfseries 52} (2020) 33} [\href{https://arxiv.org/abs/2001.10135}{{\ttfamily
  2001.10135}}].

\bibitem{Cartan:1922}
{\'E}.~Cartan, \emph{Sur une g{\'e}n{\'e}ralisation de la notion de courbure de
  riemann et les espaces {\`a} torsion}, {\emph{Comptes Rendus, Ac. Sc. Paris}
  {\bfseries 174} (1922) 593}.

\bibitem{Cartan:1923}
{\'E}.~Cartan, \emph{Sur les vari{\'e}t{\'e}s {\`a} connexion affine et la
  th{\'e}orie de la relativit{\'e} g{\'e}n{\'e}ralis{\'e}e (premi{\`e}re
  partie)},  in \emph{Annales scientifiques de l'{\'E}cole normale
  sup{\'e}rieure}, vol.~40, pp.~325--412, 1923.

\bibitem{Cartan:1924}
{\'E}.~Cartan, \emph{Sur les vari{\'e}t{\'e}s {\`a} connexion affine, et la
  th{\'e}orie de la relativit{\'e} g{\'e}n{\'e}ralis{\'e}e (premi{\`e}re
  partie)(suite)},  in \emph{Annales scientifiques de l'{\'E}cole Normale
  Sup{\'e}rieure}, vol.~41, pp.~1--25, 1924.

\bibitem{Cartan1925}
{\'E}.~Cartan, \emph{Sur les vari{\'e}t{\'e}s {\`a} connexion affine, et la
  th{\'e}orie de la relativit{\'e} g{\'e}n{\'e}ralis{\'e}e (deuxi{\`e}me
  partie)},  in \emph{Annales scientifiques de l'{\'E}cole normale
  sup{\'e}rieure}, vol.~42, pp.~17--88, 1925.

\bibitem{Einstein1928}
A.~Einstein, \emph{Riemanngeometrie mit aufrechterhaltung des begriffes des
  fern-parallelismus}, {\emph{Sitzungsber. Preuss. Akad. Wiss} (1928) 217}.

\bibitem{Einstein19282}
A.~Einstein, \emph{Neue m{\"o}glichkeit f{\"u}r eine einheitliche feldtheorie
  von gravitation und elektrizit{\"a}t}, {\emph{Sitzungsber. Preuss. Akad.
  Wiss} (1928) 224}.

\bibitem{Utiyama:1956sy}
R.~Utiyama, \emph{{Invariant theoretical interpretation of interaction}},
  \href{https://doi.org/10.1103/PhysRev.101.1597}{\emph{Phys. Rev.} {\bfseries
  101} (1956) 1597}.

\bibitem{sciama1962analogy}
D.W.~Sciama, \emph{On the analogy between charge and spin in general
  relativity}, {\emph{Recent developments in general relativity} (1962) 415}.

\bibitem{Kibble:1961ba}
T.~Kibble, \emph{{Lorentz invariance and the gravitational field}},
  \href{https://doi.org/10.1063/1.1703702}{\emph{J. Math. Phys.} {\bfseries 2}
  (1961) 212}.

\bibitem{Hehl:1976kj}
F.~Hehl, P.~Von Der~Heyde, G.~Kerlick and J.~Nester, \emph{{General Relativity
  with Spin and Torsion: Foundations and Prospects}},
  \href{https://doi.org/10.1103/RevModPhys.48.393}{\emph{Rev. Mod. Phys.}
  {\bfseries 48} (1976) 393}.

\bibitem{Shapiro:2001rz}
I.~Shapiro, \emph{{Physical aspects of the space-time torsion}},
  \href{https://doi.org/10.1016/S0370-1573(01)00030-8}{\emph{Phys. Rept.}
  {\bfseries 357} (2002) 113}
  [\href{https://arxiv.org/abs/hep-th/0103093}{{\ttfamily hep-th/0103093}}].

\bibitem{Dadhich:2010xa}
N.~Dadhich and J.M.~Pons, \emph{{On the equivalence of the Einstein-Hilbert and
  the Einstein-Palatini formulations of general relativity for an arbitrary
  connection}}, \href{https://doi.org/10.1007/s10714-012-1393-9}{\emph{Gen.
  Rel. Grav.} {\bfseries 44} (2012) 2337}
  [\href{https://arxiv.org/abs/1010.0869}{{\ttfamily 1010.0869}}].

\bibitem{Shaposhnikov:2020frq}
M.~Shaposhnikov, A.~Shkerin, I.~Timiryasov and S.~Zell, \emph{{Einstein-Cartan
  gravity, matter, and scale-invariant generalization}},
  \href{https://doi.org/10.1007/JHEP10(2020)177}{\emph{JHEP} {\bfseries 10}
  (2020) 177} [Erratum:
  \href{https://doi.org/10.1007/JHEP08(2021)162}{\textit{ibid.} \textbf{08}
  (2021) 162}], [\href{https://arxiv.org/abs/2007.16158}{{\ttfamily
  2007.16158}}].

\bibitem{Hojman:1980kv}
R.~Hojman, C.~Mukku and W.~Sayed, \emph{{Parity violation in metric torsion
  theories of gravitation}},
  \href{https://doi.org/10.1103/PhysRevD.22.1915}{\emph{Phys. Rev. D}
  {\bfseries 22} (1980) 1915}.

\bibitem{Nelson:1980ph}
P.C.~Nelson, \emph{{Gravity With Propagating Pseudoscalar Torsion}},
  \href{https://doi.org/10.1016/0375-9601(80)90348-5}{\emph{Phys. Lett. A}
  {\bfseries 79} (1980) 285}.

\bibitem{Castellani:1991et}
L.~Castellani, R.~D'Auria and P.~Fr\`e, \emph{{Supergravity and Superstrings: a
  Geometric Perspective. Vol. 1: Mathematical Foundations}} (1991).

\bibitem{Holst:1995pc}
S.~Holst, \emph{{Barbero's Hamiltonian derived from a generalized
  Hilbert-Palatini action}},
  \href{https://doi.org/10.1103/PhysRevD.53.5966}{\emph{Phys. Rev. D}
  {\bfseries 53} (1996) 5966}
  [\href{https://arxiv.org/abs/gr-qc/9511026}{{\ttfamily gr-qc/9511026}}].

\bibitem{Nieh:1981ww}
H.~Nieh and M.~Yan, \emph{{An Identity in Riemann-Cartan Geometry}},
  \href{https://doi.org/10.1063/1.525379}{\emph{J. Math. Phys.} {\bfseries 23}
  (1982) 373}.

\bibitem{Nieh:2013ada}
H.T.~Nieh, \emph{{A Torsional Topological Invariant}},  in \emph{{Conference in
  Honor of C.N. Yang's 85th Birthday}: {Statistical Physics, High Energy,
  Condensed Matter and Mathematical Physics}}, pp.~29--37, 2008
  [\href{https://arxiv.org/abs/1309.0915}{{\ttfamily 1309.0915}}].

\bibitem{Shaposhnikov:2020geh}
M.~Shaposhnikov, A.~Shkerin and S.~Zell, \emph{{Standard Model Meets Gravity:
  Electroweak Symmetry Breaking and Inflation}},
  \href{https://doi.org/10.1103/PhysRevD.103.033006}{\emph{Phys. Rev. D}
  {\bfseries 103} (2021) 033006}
  [\href{https://arxiv.org/abs/2001.09088}{{\ttfamily 2001.09088}}].

\bibitem{Bauer:2010jg}
F.~Bauer and D.A.~Demir, \emph{{Higgs-Palatini Inflation and Unitarity}},
  \href{https://doi.org/10.1016/j.physletb.2011.03.042}{\emph{Phys. Lett. B}
  {\bfseries 698} (2011) 425}
  [\href{https://arxiv.org/abs/1012.2900}{{\ttfamily 1012.2900}}].

\bibitem{0902.4465}
C.~Burgess, H.M.~Lee and M.~Trott, \emph{{Power-counting and the Validity of
  the Classical Approximation During Inflation}},
  \href{https://doi.org/10.1088/1126-6708/2009/09/103}{\emph{JHEP} {\bfseries
  09} (2009) 103} [\href{https://arxiv.org/abs/0902.4465}{{\ttfamily
  0902.4465}}].

\bibitem{0903.0355}
J.~Barbon and J.~Espinosa, \emph{{On the Naturalness of Higgs Inflation}},
  \href{https://doi.org/10.1103/PhysRevD.79.081302}{\emph{Phys. Rev. D}
  {\bfseries 79} (2009) 081302}
  [\href{https://arxiv.org/abs/0903.0355}{{\ttfamily 0903.0355}}].

\bibitem{Shaposhnikov:2020fdv}
M.~Shaposhnikov, A.~Shkerin and S.~Zell, \emph{{Quantum Effects in Palatini
  Higgs Inflation}},
  \href{https://doi.org/10.1088/1475-7516/2020/07/064}{\emph{JCAP} {\bfseries
  07} (2020) 064} [\href{https://arxiv.org/abs/2002.07105}{{\ttfamily
  2002.07105}}].

\bibitem{Shaposhnikov:2018xkv}
M.~Shaposhnikov and A.~Shkerin, \emph{{Conformal symmetry: towards the link
  between the Fermi and the Planck scales}},
  \href{https://doi.org/10.1016/j.physletb.2018.06.068}{\emph{Phys. Lett. B}
  {\bfseries 783} (2018) 253}
  [\href{https://arxiv.org/abs/1803.08907}{{\ttfamily 1803.08907}}].

\bibitem{Shaposhnikov:2018jag}
M.~Shaposhnikov and A.~Shkerin, \emph{{Gravity, Scale Invariance and the
  Hierarchy Problem}},
  \href{https://doi.org/10.1007/JHEP10(2018)024}{\emph{JHEP} {\bfseries 10}
  (2018) 024} [\href{https://arxiv.org/abs/1804.06376}{{\ttfamily
  1804.06376}}].

\bibitem{Shkerin:2019mmu}
A.~Shkerin, \emph{{Dilaton-assisted generation of the Fermi scale from the
  Planck scale}}, \href{https://doi.org/10.1103/PhysRevD.99.115018}{\emph{Phys.
  Rev. D} {\bfseries 99} (2019) 115018}
  [\href{https://arxiv.org/abs/1903.11317}{{\ttfamily 1903.11317}}].

\bibitem{Karananas:2020qkp}
G.K.~Karananas, M.~Michel and J.~Rubio, \emph{{One residue to rule them all:
  Electroweak symmetry breaking, inflation and field-space geometry}},
  \href{https://doi.org/10.1016/j.physletb.2020.135876}{\emph{Phys. Lett. B}
  {\bfseries 811} (2020) 135876}
  [\href{https://arxiv.org/abs/2006.11290}{{\ttfamily 2006.11290}}].

\bibitem{Shaposhnikov:2020aen}
M.~Shaposhnikov, A.~Shkerin, I.~Timiryasov and S.~Zell, \emph{{Einstein-Cartan
  Portal to Dark Matter}},
  \href{https://doi.org/10.1103/PhysRevLett.126.161301}{\emph{Phys. Rev. Lett.}
  {\bfseries 126} (2021) 161301} [Erratum:
  \href{https://doi.org/10.1103/PhysRevLett.127.169901}{\textit{ibid.}
  \textbf{127} (2021) 169901}],
  [\href{https://arxiv.org/abs/2008.11686}{{\ttfamily 2008.11686}}].

\bibitem{Immirzi:1996dr}
G.~Immirzi, \emph{{Quantum gravity and Regge calculus}},
  \href{https://doi.org/10.1016/S0920-5632(97)00354-X}{\emph{Nucl. Phys. B
  Proc. Suppl.} {\bfseries 57} (1997) 65}
  [\href{https://arxiv.org/abs/gr-qc/9701052}{{\ttfamily gr-qc/9701052}}].

\bibitem{Immirzi:1996di}
G.~Immirzi, \emph{{Real and complex connections for canonical gravity}},
  \href{https://doi.org/10.1088/0264-9381/14/10/002}{\emph{Class. Quant. Grav.}
  {\bfseries 14} (1997) L177}
  [\href{https://arxiv.org/abs/gr-qc/9612030}{{\ttfamily gr-qc/9612030}}].

\bibitem{hep-th/0507253}
L.~Freidel, D.~Minic and T.~Takeuchi, \emph{{Quantum gravity, torsion, parity
  violation and all that}},
  \href{https://doi.org/10.1103/PhysRevD.72.104002}{\emph{Phys. Rev.}
  {\bfseries D72} (2005) 104002}
  [\href{https://arxiv.org/abs/hep-th/0507253}{{\ttfamily hep-th/0507253}}].

\bibitem{Alexandrov:2008iy}
S.~Alexandrov, \emph{{Immirzi parameter and fermions with non-minimal
  coupling}},
  \href{https://doi.org/10.1088/0264-9381/25/14/145012}{\emph{Class. Quant.
  Grav.} {\bfseries 25} (2008) 145012}
  [\href{https://arxiv.org/abs/0802.1221}{{\ttfamily 0802.1221}}].

\bibitem{1104.2432}
D.~Diakonov, A.G.~Tumanov and A.A.~Vladimirov, \emph{{Low-energy General
  Relativity with torsion: A Systematic derivative expansion}},
  \href{https://doi.org/10.1103/PhysRevD.84.124042}{\emph{Phys. Rev.}
  {\bfseries D84} (2011) 124042}
  [\href{https://arxiv.org/abs/1104.2432}{{\ttfamily 1104.2432}}].

\bibitem{1212.0585}
J.~Magueijo, T.G.~Zlosnik and T.W.B.~Kibble, \emph{{Cosmology with a spin}},
  \href{https://doi.org/10.1103/PhysRevD.87.063504}{\emph{Phys. Rev.}
  {\bfseries D87} (2013) 063504}
  [\href{https://arxiv.org/abs/1212.0585}{{\ttfamily 1212.0585}}].

\bibitem{Rubio:2019ypq}
J.~Rubio and E.S.~Tomberg, \emph{{Preheating in Palatini Higgs inflation}},
  \href{https://doi.org/10.1088/1475-7516/2019/04/021}{\emph{JCAP} {\bfseries
  04} (2019) 021} [\href{https://arxiv.org/abs/1902.10148}{{\ttfamily
  1902.10148}}].

\bibitem{Ema:2016dny}
Y.~Ema, R.~Jinno, K.~Mukaida and K.~Nakayama, \emph{{Violent Preheating in
  Inflation with Nonminimal Coupling}},
  \href{https://doi.org/10.1088/1475-7516/2017/02/045}{\emph{JCAP} {\bfseries
  02} (2017) 045} [\href{https://arxiv.org/abs/1609.05209}{{\ttfamily
  1609.05209}}].

\bibitem{DeCross:2016cbs}
M.P.~DeCross, D.I.~Kaiser, A.~Prabhu, C.~Prescod-Weinstein and
  E.I.~Sfakianakis, \emph{{Preheating after multifield inflation with
  nonminimal couplings, III: Dynamical spacetime results}},
  \href{https://doi.org/10.1103/PhysRevD.97.023528}{\emph{Phys. Rev. D}
  {\bfseries 97} (2018) 023528}
  [\href{https://arxiv.org/abs/1610.08916}{{\ttfamily 1610.08916}}].

\bibitem{Tanabashi:2018oca}
{\scshape Particle Data Group} collaboration, \emph{{Review of Particle
  Physics}}, \href{https://doi.org/10.1103/PhysRevD.98.030001}{\emph{Phys. Rev.
  D} {\bfseries 98} (2018) 030001}.

\bibitem{Bezrukov:2014ina}
F.~Bezrukov and M.~Shaposhnikov, \emph{{Why Should We Care About the Top Quark
  Yukawa Coupling?}}, \href{https://doi.org/10.1134/S1063776115030152}{\emph{J.
  Exp. Theor. Phys.} {\bfseries 120} (2015) 335}
  [\href{https://arxiv.org/abs/1411.1923}{{\ttfamily 1411.1923}}].

\bibitem{Sirunyan:2019zvx}
{\scshape CMS} collaboration, \emph{{Measurement of $\mathrm{t\bar t}$
  normalised multi-differential cross sections in pp collisions at $\sqrt s=13$
  TeV, and simultaneous determination of the strong coupling strength, top
  quark pole mass, and parton distribution functions}},
  \href{https://doi.org/10.1140/epjc/s10052-020-7917-7}{\emph{Eur. Phys. J. C}
  {\bfseries 80} (2020) 658}
  [\href{https://arxiv.org/abs/1904.05237}{{\ttfamily 1904.05237}}].

\bibitem{Aad:2019mkw}
{\scshape ATLAS} collaboration, \emph{{Measurement of the top-quark mass in
  $t\bar{t}+1$-jet events collected with the ATLAS detector in $pp$ collisions
  at $\sqrt{s}=8$ TeV}},
  \href{https://doi.org/10.1007/JHEP11(2019)150}{\emph{JHEP} {\bfseries 11}
  (2019) 150} [\href{https://arxiv.org/abs/1905.02302}{{\ttfamily
  1905.02302}}].

\bibitem{Bezrukov:2014ipa}
F.~Bezrukov, J.~Rubio and M.~Shaposhnikov, \emph{{Living beyond the edge: Higgs
  inflation and vacuum metastability}},
  \href{https://doi.org/10.1103/PhysRevD.92.083512}{\emph{Phys. Rev. D}
  {\bfseries 92} (2015) 083512}
  [\href{https://arxiv.org/abs/1412.3811}{{\ttfamily 1412.3811}}].

\bibitem{Bezrukov:2010jz}
F.~Bezrukov, A.~Magnin, M.~Shaposhnikov and S.~Sibiryakov, \emph{{Higgs
  inflation: consistency and generalisations}},
  \href{https://doi.org/10.1007/JHEP01(2011)016}{\emph{JHEP} {\bfseries 01}
  (2011) 016} [\href{https://arxiv.org/abs/1008.5157}{{\ttfamily 1008.5157}}].

\bibitem{1012.2900}
F.~Bauer and D.A.~Demir, \emph{{Higgs-Palatini Inflation and Unitarity}},
  \href{https://doi.org/10.1016/j.physletb.2011.03.042}{\emph{Phys. Lett. B}
  {\bfseries 698} (2011) 425}
  [\href{https://arxiv.org/abs/1012.2900}{{\ttfamily 1012.2900}}].

\bibitem{Shaposhnikov:2008xb}
M.~Shaposhnikov and D.~Zenhausern, \emph{{Scale invariance, unimodular gravity
  and dark energy}},
  \href{https://doi.org/10.1016/j.physletb.2008.11.054}{\emph{Phys. Lett. B}
  {\bfseries 671} (2009) 187}
  [\href{https://arxiv.org/abs/0809.3395}{{\ttfamily 0809.3395}}].

\bibitem{GarciaBellido:2011de}
J.~Garcia-Bellido, J.~Rubio, M.~Shaposhnikov and D.~Zenhausern,
  \emph{{Higgs-Dilaton Cosmology: From the Early to the Late Universe}},
  \href{https://doi.org/10.1103/PhysRevD.84.123504}{\emph{Phys. Rev. D}
  {\bfseries 84} (2011) 123504}
  [\href{https://arxiv.org/abs/1107.2163}{{\ttfamily 1107.2163}}].

\bibitem{Bezrukov:2012hx}
F.~Bezrukov, G.K.~Karananas, J.~Rubio and M.~Shaposhnikov, \emph{{Higgs-Dilaton
  Cosmology: an effective field theory approach}},
  \href{https://doi.org/10.1103/PhysRevD.87.096001}{\emph{Phys. Rev. D}
  {\bfseries 87} (2013) 096001}
  [\href{https://arxiv.org/abs/1212.4148}{{\ttfamily 1212.4148}}].

\bibitem{Rubio:2020zht}
J.~Rubio, \emph{{Scale symmetry, the Higgs and the Cosmos}},
  \href{https://doi.org/10.22323/1.376.0074}{\emph{PoS} {\bfseries CORFU2019}
  (2020) 074} [\href{https://arxiv.org/abs/2004.00039}{{\ttfamily
  2004.00039}}].

\bibitem{ForemanMackey:2012ig}
D.~Foreman-Mackey, D.W.~Hogg, D.~Lang and J.~Goodman, \emph{{emcee: The MCMC
  Hammer}}, \href{https://doi.org/10.1086/670067}{\emph{Publ. Astron. Soc.
  Pac.} {\bfseries 125} (2013) 306}
  [\href{https://arxiv.org/abs/1202.3665}{{\ttfamily 1202.3665}}].

\bibitem{Langvik:2020nrs}
M.~L\r{a}ngvik, J.-M.~Ojanper\"a, S.~Raatikainen and S.~R\"as\"anen,
  \emph{{Higgs inflation with the Holst and the Nieh\textendash{}Yan term}},
  \href{https://doi.org/10.1103/PhysRevD.103.083514}{\emph{Phys. Rev. D}
  {\bfseries 103} (2021) 083514}
  [\href{https://arxiv.org/abs/2007.12595}{{\ttfamily 2007.12595}}].

\end{thebibliography}\endgroup

\end{document}